\documentclass[twocolumn,reprint,amsmath,superscriptaddress,amssymb,aps]{revtex4-1}\usepackage[utf8]{inputenc}
\usepackage{graphicx}
\usepackage{dcolumn}
\usepackage{comment}

\usepackage{braket}
\usepackage{bbold}
\usepackage{xcolor}
\usepackage{color,soul}
\usepackage{graphicx}

\begin{document}
\title{Nuclear spin-wave quantum register for a solid state qubit}
\author{Andrei~Ruskuc}
\affiliation{Thomas J. Watson, Sr, Laboratory of Applied Physics, California Institute of Technology, Pasadena, CA, USA}
\affiliation{Kavli Nanoscience Institute, California Institute of Technology, Pasadena, CA, USA}
\affiliation{Institute for Quantum Information and Matter, California Institute of Technology, Pasadena, CA, USA}
\author{Chun-Ju~Wu}
\affiliation{Thomas J. Watson, Sr, Laboratory of Applied Physics, California Institute of Technology, Pasadena, CA, USA}
\affiliation{Kavli Nanoscience Institute, California Institute of Technology, Pasadena, CA, USA}
\affiliation{Institute for Quantum Information and Matter, California Institute of Technology, Pasadena, CA, USA}
\affiliation{Division of Physics, Mathematics and Astronomy, California Institute of Technology, Pasadena, CA, USA}
\author{Jake~Rochman}
\affiliation{Thomas J. Watson, Sr, Laboratory of Applied Physics, California Institute of Technology, Pasadena, CA, USA}
\affiliation{Kavli Nanoscience Institute, California Institute of Technology, Pasadena, CA, USA}
\affiliation{Institute for Quantum Information and Matter, California Institute of Technology, Pasadena, CA, USA}
\author{Joonhee~Choi}
\email{joonhee@caltech.edu}
\affiliation{Institute for Quantum Information and Matter, California Institute of Technology, Pasadena, CA, USA}
\affiliation{Division of Physics, Mathematics and Astronomy, California Institute of Technology, Pasadena, CA, USA}
\author{Andrei~Faraon}
\email{faraon@caltech.edu}
\affiliation{Thomas J. Watson, Sr, Laboratory of Applied Physics, California Institute of Technology, Pasadena, CA, USA}
\affiliation{Kavli Nanoscience Institute, California Institute of Technology, Pasadena, CA, USA}
\affiliation{Institute for Quantum Information and Matter, California Institute of Technology, Pasadena, CA, USA}
\maketitle
\renewcommand{\figurename}{Fig.}
\textbf{Solid-state nuclear spins surrounding individual, optically addressable qubits \cite{Awschalom2018,Chatterjee2021} provide a crucial resource for quantum networks \cite{Briegel1998,Hensen2015,Bhaskar2020,Pompili2021}, computation \cite{Waldherr2014,Taminiau2014,Zhong2019,Bradley2019a,Kinos2021} and simulation \cite{Randall2021}. While hosts with sparse nuclear spin baths are typically chosen to mitigate qubit decoherence \cite{Wolfowicz2021}, developing coherent quantum systems in nuclear spin-rich hosts enables exploration of a much broader range of materials for quantum information applications. The collective modes of these dense nuclear spin ensembles provide a natural basis for quantum storage \cite{Taylor2003}, however, utilizing them as a resource for single spin qubits has thus far remained elusive. Here, by using a highly coherent, optically addressed \textsuperscript{171}Yb\textsuperscript{3+} qubit doped into a nuclear spin-rich yttrium orthovanadate crystal \cite{Kindem2020}, we develop a robust quantum control protocol to manipulate the multi-level nuclear spin states of neighbouring \textsuperscript{51}V\textsuperscript{5+} lattice ions. Via a dynamically-engineered spin exchange interaction, we polarise this nuclear spin ensemble, generate collective spin excitations, and subsequently use them to implement a long-lived quantum memory. We additionally demonstrate preparation and measurement of maximally entangled \textsuperscript{171}Yb--\textsuperscript{51}V Bell states. Unlike conventional, disordered nuclear spin based quantum memories \cite{GurudevDutt2007,Kolkowitz2012,Taminiau2012,Zhao2012,Metsch2019,Bourassa2020a,Hensen2020,Kornher2020,Wolfowicz2016}, our platform is deterministic and reproducible, ensuring identical quantum registers for all \textsuperscript{171}Yb\textsuperscript{3+} qubits. Our approach provides a framework for utilising the complex structure of dense nuclear spin baths, paving the way for building large-scale quantum networks using single rare-earth ion qubits \cite{Utikal2014,Siyushev2014,Zhong2018,Chen2020,Kindem2020}.}

We recently demonstrated that at zero magnetic field, the hyperfine levels of single \textsuperscript{171}Yb\textsuperscript{3+} ions doped into yttrium orthovanadate (YVO\textsubscript{4}), coupled to nanophotonic cavities, form high-quality optically addressable qubits \cite{Kindem2020} (Fig.~1a). The surrounding \textsuperscript{51}V\textsuperscript{5+} lattice ion nuclear spins generate a noisy magnetic field environment due to their large magnetic moment and high spin (I=7/2). Coherent \textsuperscript{171}Yb qubit operation is enabled by magnetically-insensitive transitions, leading to long coherence times (16~ms) and high gate fidelities (0.99975) (Extended~Data~Fig.~2). Whilst decoupling from sources of magnetic noise achieves an excellent operating regime for the \textsuperscript{171}Yb qubit, the \textsuperscript{51}V nuclear spins also provide a readily accessible, local resource for quantum information storage due to their inherently weak interactions with the environment. To date, most research regarding host nuclear spin utilisation has focused on several spectrally distinguishable impurity nuclear spins coupled to a localised electronic spin, e.g. \textsuperscript{13}C coupled to colour centres in diamond or \textsuperscript{29}Si coupled to defects in silicon carbide, rare-earth ions, quantum dots or donor qubits in silicon \cite{GurudevDutt2007,Kolkowitz2012,Taminiau2012,Zhao2012,Bradley2019a,Metsch2019,Bourassa2020a,Hensen2020,Kornher2020,Wolfowicz2016}. Recently, a regime consisting of a large number of indistinguishable nuclear spins coupled to the delocalised electronic spin in a quantum dot has also been explored \cite{Gangloff2019,Gangloff2020}. In contrast, our system addresses a new regime where a small, deterministic cluster of spectrally {\it indistinguishable} nuclear spins are coupled to a single {\it localized} electronic spin. Specifically, the \textsuperscript{171}Yb electronic wavefunction is confined to the lattice site, and the YVO\textsubscript{4} crystal consists of highly isotopically pure nuclear spins (99.8\% \textsuperscript{51}V). This confined, dense nuclear spin ensemble could be used as a deterministic local quantum processor by creating and manipulating entangled states, such as collective spin wave-like excitations, for near-term quantum applications. Critically, interfacing with these nuclear spins whilst preserving high qubit coherence necessitates the development of novel quantum control protocols using magnetically insensitive transitions that are robust against environmental noise.

\begin{center}
\begin{figure*}
\includegraphics[width=0.96\textwidth]{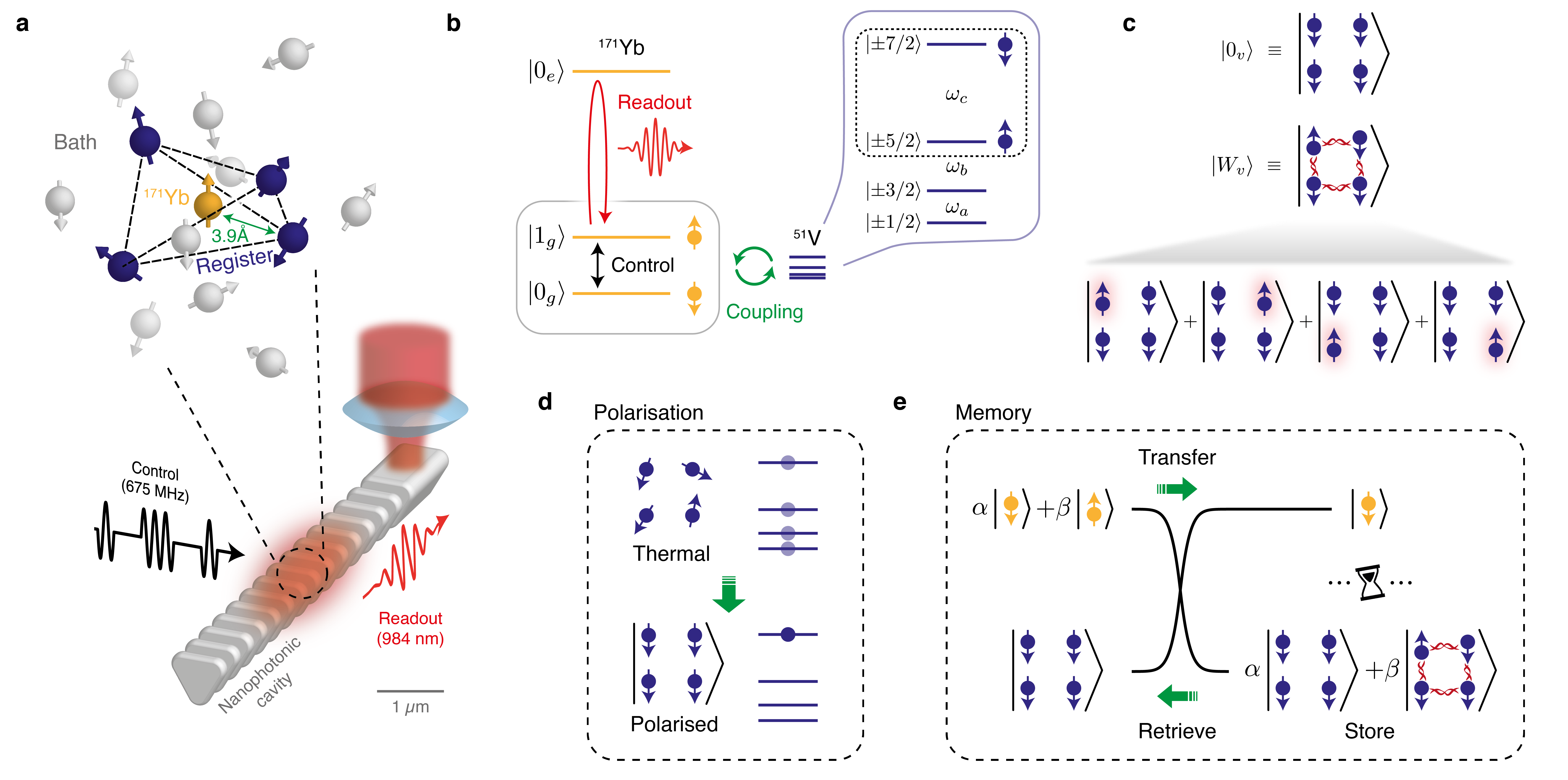}
\caption{{\bf Schematic of a many-body nuclear spin register for optically-coupled \textsuperscript{171}Yb qubits in nanophotonic cavities}. {\bf a,} Optically addressable \textsuperscript{171}Yb ion (yellow) surrounded by a local ensemble of nuclear spins from lattice \textsuperscript{51}V ions. The register (blue) consists of four \textsuperscript{51}V spins equidistantly spaced by $3.9~\AA$ from the central \textsuperscript{171}Yb. The nuclear spin bath (grey) creates random magnetic noise termed the nuclear Overhauser field. A nanophotonic cavity enables optical initialisation and readout of the \textsuperscript{171}Yb ion via single-photon detection at 984~nm~ \cite{Kindem2020}. 675 MHz microwave pulses provide high-fidelity control of the \textsuperscript{171}Yb spin state. {\bf b,} Energy level structure of \textsuperscript{171}Yb and \textsuperscript{51}V ions. Pulse-based control of the \textsuperscript{171}Yb ground-state transition ($\ket{0_g}\leftrightarrow\ket{1_g}$) enables engineered spin-exchange interactions with neighbouring \textsuperscript{51}V ions. The energy level structure of the spin-7/2 \textsuperscript{51}V consists of four quadratically-spaced, doubly degenerate energy levels, $\{\ket{\pm m_I}\}=\{\ket{\pm1/2}$,  $\ket{\pm3/2}$,  $\ket{\pm5/2}$,  $\ket{\pm7/2}\}$, resulting in three distinct transitions, $\omega_{a,b,c}/2\pi =$ 330 kHz, 660 kHz, and 991 kHz, respectively. The $\omega_c$ transition (dotted box) is used to implement the local nuclear spin register for quantum information storage. {\bf c,} Effective qubit states of the nuclear spin register. The $\ket{0_v}$ and $\ket{W_v}$ states consist of all four \textsuperscript{51}V ions prepared in the $\ket{\downarrow}=\ket{\pm7/2}$ state and a single spin excitation equally delocalised  in the $\ket{\uparrow}=\ket{\pm5/2}$ state, respectively. {\bf d,} Initialisation of the nuclear spins from a thermal state into the polarised $\ket{\downarrow \downarrow \downarrow \downarrow}$ state. {\bf e,} Transfer of a quantum state from \textsuperscript{171}Yb to the \textsuperscript{51}V register, storage and subsequent retrieval. Both the state initialization and transfer are enabled by robust, dynamically engineered interactions between \textsuperscript{171}Yb and \textsuperscript{51}V ions.}
\end{figure*}
\end{center}

At zero-magnetic field the \textsuperscript{171}Yb ground state contains a pair of levels, $\ket{0_g}$ and $\ket{1_g}$, separated by 675 MHz, which form our qubit \cite{Kindem2018a} (Fig.~1b). We can optically read out the $\ket{1_g}$ population via a series of $\pi$ pulses at 984~nm, each followed by time-resolved detection of resonant photon emission (Extended~Data~Fig.~1). This is enabled by coupling the \textsuperscript{171}Yb ion to a nanophotonic cavity leading to high transition cyclicity, reduced optical lifetime and high photon collection efficiency \cite{Kindem2020}. The local crystalline environment consists of \textsuperscript{89}Y, \textsuperscript{51}V and \textsuperscript{16}O ions. Of these, \textsuperscript{51}V with nuclear spin 7/2 has the largest magnetic dipole moment and zero-field structure due to a quadrupole interaction with the lattice electric field \cite{Bleaney1982}. This leads to four quadratically-spaced, doubly degenerate energy levels, $\{\ket{\pm m_I}\}=$\{$\ket{\pm1/2}$,  $\ket{\pm3/2}$,  $\ket{\pm5/2}$,  $\ket{\pm7/2}$\}, and three magnetic-dipole allowed transitions between these levels $\omega_a$, $\omega_b$, $\omega_c$ (Fig.~1b).

Local \textsuperscript{51}V ions are categorised into two complementary ensembles: the register and the bath. The register spins fulfil two conditions: (1) they are constituents of the frozen core: a set of \textsuperscript{51}V ions spectrally distinguished from the bath due to proximity to \textsuperscript{171}Yb; (2) the \textsuperscript{171}Yb--\textsuperscript{51}V interaction Hamiltonian can drive transitions between their quadrupole levels. As shown later, experimental evidence suggests that the register consists of four \textsuperscript{51}V spins, equidistant from the central \textsuperscript{171}Yb (Fig.~1a). At zero field, the \textsuperscript{171}Yb $\ket{0_g}$, $\ket{1_g}$ states have no intrinsic magnetic dipole moment and thus interactions with \textsuperscript{51}V register spins are forbidden to first order. However, a weak \textsuperscript{171}Yb dipole moment is induced by a random magnetic field originating from the bath (the nuclear Overhauser field, with $z$ component $B^\text{OH}_z$), giving rise to an effective \textsuperscript{171}Yb--\textsuperscript{51}V register interaction. Specifically, a second-order perturbation analysis yields the following Hamiltonian:
\begin{equation}
\label{ddint}
\hat{H}_\text{int} =\hat{\Tilde{S}}_zB^\text{OH}_z\sum_{i\in\text{register}}\left(a_x \hat{I}_x^{(i)} +a_z \hat{I}_z^{(i)}\right),
\end{equation}
where $\hat{\Tilde{S}}_z$ is the \textsuperscript{171}Yb qubit operator along the $z$ axis in a weakly perturbed basis, $\hat{I}_{x,z}^{(i)}$ are the nuclear spin-7/2 operators along the $x, z$ axes, and $a_{x,z}$ are the coupling coefficients (Supplementary Information). Note that $B^\text{OH}_z$ varies randomly in time as the bath changes state in a stochastic fashion, rendering this interaction Hamiltonian unreliable for register quantum state manipulation. To this end, we develop a protocol to generate a {\it deterministic} \textsuperscript{171}Yb--\textsuperscript{51}V interaction via Hamiltonian engineering, which will be elaborated later.

An additional challenge is presented by the spectral indistinguishability of the register spins, necessitating storage in collective states. As originally proposed for quantum dots \cite{Taylor2003}, single spin excitations of a polarised nuclear spin ensemble can be used for quantum information storage. These states are often termed spin waves or nuclear magnons and are generated by spin-preserving exchange dynamics. Specifically, preparing these collective nuclear spin states relies firstly on initialising the thermal register ensemble into a pure state, $\ket{0_v} = \ket{\downarrow \downarrow \downarrow \downarrow}$, where $\{\ket{\uparrow},\ket{\downarrow}\}=\{\ket{\pm5/2},\ket{\pm7/2}\}$  is a two-level sub-manifold of the nuclear spin-7/2 \textsuperscript{51}V ion (Fig.~1c,d). Next, with access to exchange dynamics and \textsuperscript{171}Yb initialised in $\ket{1_g}$, we can transfer a single excitation from the \textsuperscript{171}Yb to the register. We note that the excitation is delocalised equally across the four register spins due to coupling homogeneity as determined by the lattice geometry, thus naturally realising the entangled four-body W-state $\ket{W_v}$ \cite{Weimer2013} given by
\begin{align}
 \ket{W_v}  = \frac{ \ket{\uparrow \downarrow \downarrow \downarrow } + \ket{\downarrow \uparrow \downarrow \downarrow } + \ket{\downarrow \downarrow \uparrow \downarrow } + \ket{\downarrow \downarrow \downarrow \uparrow } 
}{2}
\end{align}
(Fig.~1c). If the \textsuperscript{171}Yb qubit is initialised into $\ket{0_g}$ there are no spin excitations in the system and the \textsuperscript{51}V register remains in $\ket{0_v}$. Crucially, these dynamics realise a quantum swap gate between a target state prepared by the \textsuperscript{171}Yb qubit, $\ket{\psi} = \alpha \ket{0_g} + \beta \ket{1_g}$, and the $\ket{0_v}$ state of the \textsuperscript{51}V register, leading to
\begin{equation}
\left(\alpha\ket{0_g}+\beta\ket{1_g}\right)\ket{0_v}\rightarrow\ket{0_g}\left(\alpha\ket{0_v}+\beta\ket{W_v}\right).
\end{equation}
After waiting for a certain period of time, the stored quantum state can be retrieved from the \textsuperscript{51}V register by applying a second swap gate (Fig.~1e). Note that the spin-wave like state $\ket{W_v}$ of the nuclear ensemble is being utilized as a constituent of the quantum memory basis.

To realise this storage protocol we require \textsuperscript{171}Yb--\textsuperscript{51}V spin-exchange interactions that are independent from the random, bath-induced dipole moment (equation~\eqref{ddint}). We note that established pulse-based methods used to generate such interactions, e.g. Hartmann Hahn \cite{Hartmann1962} and PulsePol \cite{Schwartz2018}, do not suit our requirements as they are susceptible to random noise from the bath (Extended~Data~Fig.~3 and Supplementary Information). To this end, we employ a framework for robust dynamic Hamiltonian engineering \cite{Choi2019} to design a new sequence tailored for qubits with no intrinsic magnetic moment (subsequently referred to as ZenPol for `\textbf{ze}ro first-order Zeeman \textbf{n}uclear-spin \textbf{pol}arisation'). ZenPol comprises equidistant $\pi/2$ and $\pi$ pulses combined with a synchronous, $z$-directed, square-wave RF magnetic field with tuneable amplitude, $B^\text{RF}$, and period $2\tau$ (Fig.~2a). The sequence is repeated $M$ times leading to a total interrogation duration of $t_M=2\tau M$. The RF field induces an alternating \textsuperscript{171}Yb magnetic dipole moment, thereby generating a similar \textsuperscript{171}Yb--\textsuperscript{51}V interaction as $B^\text{OH}_z$ in equation~\eqref{ddint} but in a {\it controlled} manner. The sequence is synchronised with the \textsuperscript{51}V precession at one of the nuclear spin transition frequencies, $\omega_j$, by satisfying
\begin{equation}
\frac{1}{2\tau}=\frac{\omega_j}{2\pi k},
\end{equation}
with $k$ an odd integer (Extended~Data~Fig.~4). At this resonance condition the leading-order dynamics are understood by considering the temporal interference between time-varying \textsuperscript{171}Yb spin operators and \textsuperscript{51}V precession in the interaction picture (Methods). The ZenPol sequence is designed such that RF-induced spin-preserving dynamics interfere constructively, while all other dynamics, including the bath-induced incoherent interactions, undergo destructive interference. As a result, the \textsuperscript{171}Yb--\textsuperscript{51}V interaction is governed by the following time-averaged effective Hamiltonian
\begin{equation}
	\hat{H}_\text{avg} =   b_{(k,\omega_j)} B^\text{RF} \sum_{i \in \text{register}}\left( \hat{\tilde{S}}_+ \hat{\tilde{I}}^{(i)}_- + \hat{\tilde{S}}_-\hat{\tilde{I}}^{(i)}_+\right),
\end{equation}
where $b_{(k,\omega_j)}$ is a $k$-dependent prefactor for the $\omega_j$ transition, $\Hat{\tilde{I}}_+ = \ket{\uparrow} \bra{\downarrow}, \Hat{\tilde{I}}_- = \ket{\downarrow} \bra{\uparrow}$ are the raising and lowering operators in an effective nuclear two-level manifold and $\Hat{\tilde{S}}_\pm$ are similarly defined for the \textsuperscript{171}Yb qubit (Methods). We note that while the nuclear spin can stochastically occupy either the $\{\ket{+m_I}\}$ or $\{\ket{-m_I}\}$ manifold of states, our protocol is insensitive to this sign. We emphasize that the ZenPol sequence operates at zero magnetic field where a long \textsuperscript{171}Yb coherence time can be maintained; it is insensitive to the presence of random noise from the bath; and is also robust to experimental imperfections, e.g. pulse rotation errors (Methods).

We use the ZenPol sequence to perform spectroscopy of the \textsuperscript{171}Yb nuclear spin environment. Figure~2b shows a ZenPol spectrum obtained by initialising the \textsuperscript{171}Yb into $\ket{0_g}$, applying an $M=30$ period ZenPol sequence with variable inter-pulse spacing ($\tau/4$) and reading out the \textsuperscript{171}Yb population. As a result of the engineered exchange interaction, we find that the $\ket{0_g}$  population decreases significantly at expected $\tau$ values corresponding to the odd-$k$ \textsuperscript{51}V resonances (red line, Fig.~2b). Even-$k$ resonances are also observed even in the absence of the RF field, which are attributed to the incoherent interaction dominated by the random nuclear Overhauser field (blue line, Fig.~2b).

In particular, we note that all the odd-$k$ resonances are split near each isolated \textsuperscript{51}V transition (dotted boxes, Fig.~2b). For example, resonance frequencies of $\{$660~kHz, 685~kHz$\}$ and $\{$991~kHz, 1028~kHz$\}$ are identified around the $\omega_b~(k=3)$ and $\omega_c~(k=5)$ transitions, respectively. In both cases, the higher-frequency resonance agrees well with literature values extracted from NMR on YVO\textsubscript{4} crystals (685~kHz, 1027~kHz) \cite{Bleaney1982}. We therefore postulate the presence of two nuclear spin ensembles: a distant large ensemble with unperturbed frequency (constituents of the bath) and a local small ensemble with a frequency shift due to crystalline strain in the vicinity of the \textsuperscript{171}Yb ion (the register).

\begin{center}
\begin{figure*}
\includegraphics[width=0.96\textwidth]{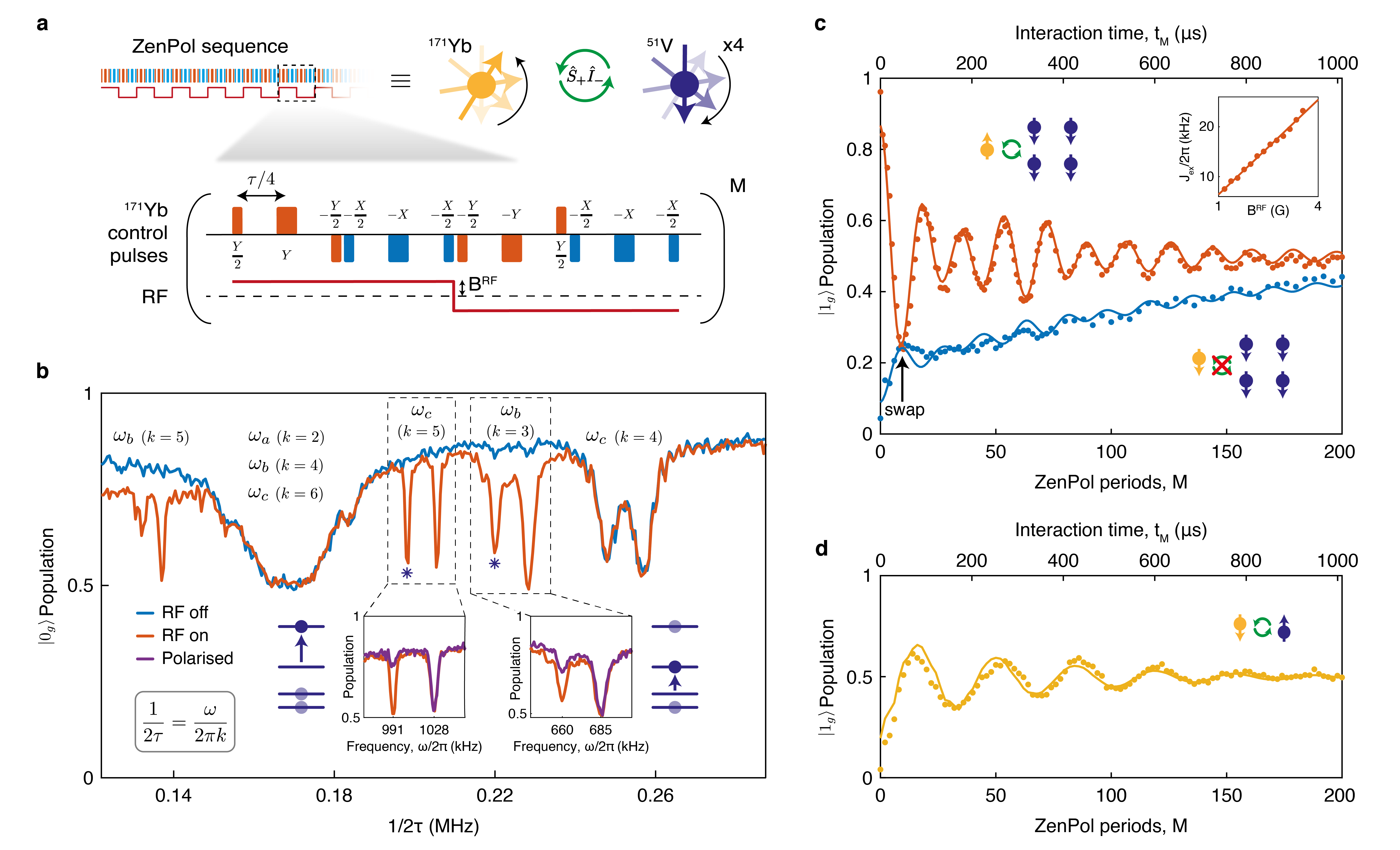}
\caption{
{\bf Pulse-based Hamiltonian engineering, nuclear register polarisation and spin exchange between \textsuperscript{171}Yb and \textsuperscript{51}V ions.} {\bf a,} Engineered spin-exchange interactions via our ZenPol sequence. Equidistant $\pi/2$ and $\pi$ pulses combined with a square-wave RF magnetic field with amplitude $B^\text{RF}$ are periodically applied to the \textsuperscript{171}Yb qubit with base sequence period $2\tau$. {\bf b,} ZenPol sequence spectroscopy. \textsuperscript{171}Yb--\textsuperscript{51}V resonance is achieved for a given \textsuperscript{51}V transition, $\omega_j$, when $1/2\tau = \omega_j/2\pi k$ with integer $k$. We use the isolated, RF-induced $\omega_c\ (k=5)$ and $\omega_b\ (k=3)$ transitions to polarize the multi-level nuclear spins of neighbouring \textsuperscript{51}V ions (dashed boxes). Both cases exhibit split-resonance features, attributed to the presence of two distinct \textsuperscript{51}V ensembles: the four \textsuperscript{51}V register spins (starred transitions) adjacent to the \textsuperscript{171}Yb qubit experience a frozen-core type detuning relative to the more distant bath. Insets: Under repeated application of the ZenPol sequence targeted at the $\omega_b$ or $\omega_c$ register transitions and  interleaved with \textsuperscript{171}Yb initialisation, the four register spins are selectively polarised (purple lines). {\bf c,} Spin-exchange dynamics with the four \textsuperscript{51}V register spins. The \textsuperscript{171}Yb qubit and \textsuperscript{51}V register spins are initialized into $\ket{1_g}$ and $\ket{0_v} (\equiv\ket{\downarrow\downarrow\downarrow\downarrow})$, respectively. Subsequently, our pulse sequence induces resonant spin exchange on the $\omega_c$ transition leading to oscillation between $\ket{1_g}\ket{0_v} \leftrightarrow \ket{0_g}\ket{W_v}$ where $\ket{W_v}$ is a spin-wave like W-state (red markers). Inset: the rate of spin exchange scales linearly with $B^\text{RF}$. With \textsuperscript{171}Yb in $\ket{0_g}$, there are no spin excitations in the system and oscillations are suppressed (blue markers). We use a ZenPol sequence with $M=10$ periods and duration $t_M=50~\mu$s to realise a swap gate (black arrow).  {\bf d,} Spin-exchange dynamics with a single \textsuperscript{51}V nuclear spin. Three \textsuperscript{51}V spins are shelved in $\ket{\pm3/2}$ and a single spin is excited to $\ket{\uparrow}=\ket{\pm5/2}$. Accordingly, the \textsuperscript{171}Yb qubit undergoes spin exchange with the $\omega_c$ transition manifold at a reduced oscillation frequency. In c,d, solid lines are from simulations with phenomenological exponential decay constants (Supplementary Information).
}
\end{figure*}
\end{center}

Polarisation of the entire nuclear spin register relies on repeated application of the ZenPol sequence, resonant with a targeted transition, interleaved with reinitialisation of the \textsuperscript{171}Yb qubit leading to unidirectional transfer of \textsuperscript{51}V population (Extended~Data~Fig.~1c). Since the spin-7/2 \textsuperscript{51}V ions have four doubly-degenerate energy levels, we achieve high fidelity initialisation by independently polarising different transitions with different values of $\tau$. For example, to prepare the register spins in $\ket{\pm7/2}=\ket{\downarrow}$, we repeatedly apply a pair of ZenPol sequences which first polarise into $\ket{\pm5/2}$ using the $\omega_b$ transition, and then subsequently into $\ket{\pm7/2}$ using the $\omega_c$ transition (Extended~Data~Fig.~5). We confirm that both $\omega_b$ and $\omega_c$ transitions of the \textsuperscript{51}V register are successfully polarised as indicated by the near-complete disappearance of the initial resonances (insets, Fig.~2b). Note that the resonances at 685~kHz and 1028~kHz are unaffected, corroborating our speculation on the existence of two distinct \textsuperscript{51}V ensembles discussed above. The $\omega_a$ transition is not directly addressed by the ZenPol sequence due to spectral overlap with other resonances, however, this does not limit our polarisation fidelity, estimated to be $\approx84\%$, as discussed in Supplementary Information.

After initialising all four register \textsuperscript{51}V spins into a polarized state $\ket{0_v} = \ket{\downarrow \downarrow \downarrow \downarrow}$, the ZenPol sequence can also induce coherent oscillations of a single spin excitation between the \textsuperscript{171}Yb ion and the polarised \textsuperscript{51}V ensemble. Figure~2c shows the \textsuperscript{171}Yb population as a function of sequence period, $M$, when the single-spin exchange is targeted at the $\omega_c$ transition. With \textsuperscript{171}Yb initialised in $\ket{1_g}$, the quantum state evolves according to: 
\begin{multline}
\label{magnonosc}
\ket{\psi(t_M)} = \ket{1_g} \ket{0_v} \cos(J_\text{ex} t_M/2)\\
-i \ket{0_g} \ket{W_v} \sin(J_\text{ex} t_M/2)
\end{multline}
with spin-exchange rate $J_\text{ex}=4b_{(5,\omega_c)}B^\text{RF}$ (red, Fig.~2c). Note that when $J_\text{ex}t_M = \pi$, the sequence realises a swap gate (black arrow, Fig.~2c), whereby a single-spin excitation is completely transferred to the register, i.e., $\ket{1_g}\ket{0_v} \rightarrow \ket{0_g}\ket{W_v}$. Furthermore, we emphasize that $J_\text{ex}$ can be accurately controlled by varying $B^\text{RF}$, allowing for fidelity optimisation of the swap gate (inset, Fig.~2c). By contrast, with \textsuperscript{171}Yb initialised in $\ket{0_g}$, exchange interactions are forbidden and thus oscillations are suppressed (blue, Fig.~2c).

We note that the spin-exchange rate is collectively enhanced by a factor of $\sqrt{N}$, where $N$ is the number of indistinguishable spins forming the register. We verify this by controlling the number of spins in the $\omega_c$ transition manifold and measuring the effect on $J_\text{ex}$. This is implemented by first emptying the $\omega_c$ manifold via the application of downward-polarising ZenPol sequences, thereby pumping all four spins to $\ket{\pm3/2}$ and $\ket{\pm1/2}$. Subsequently, a single excitation is performed on the $\omega_b$ transition to flip one spin from $\ket{\pm3/2}$ to $\ket{\uparrow} (= \ket{\pm5/2})$, leading to $N=1$ spins in the $\omega_c$ manifold (Supplementary Information). Applying a ZenPol sequence resonant with the $\omega_c$ transition, we find that the resulting exchange frequency is reduced by a factor of $\approx \sqrt{4}$ (Fig.~2d); according to the YVO\textsubscript{4} lattice structure, the register likely consists of the second-nearest shell of four equidistant \textsuperscript{51}V ions (Supplementary Information). This assumption is supported by close agreement between experiment and numerical simulation in all cases (Extended~Data~Fig.~6).

\begin{center}
\begin{figure}
\includegraphics[width=0.48\textwidth]{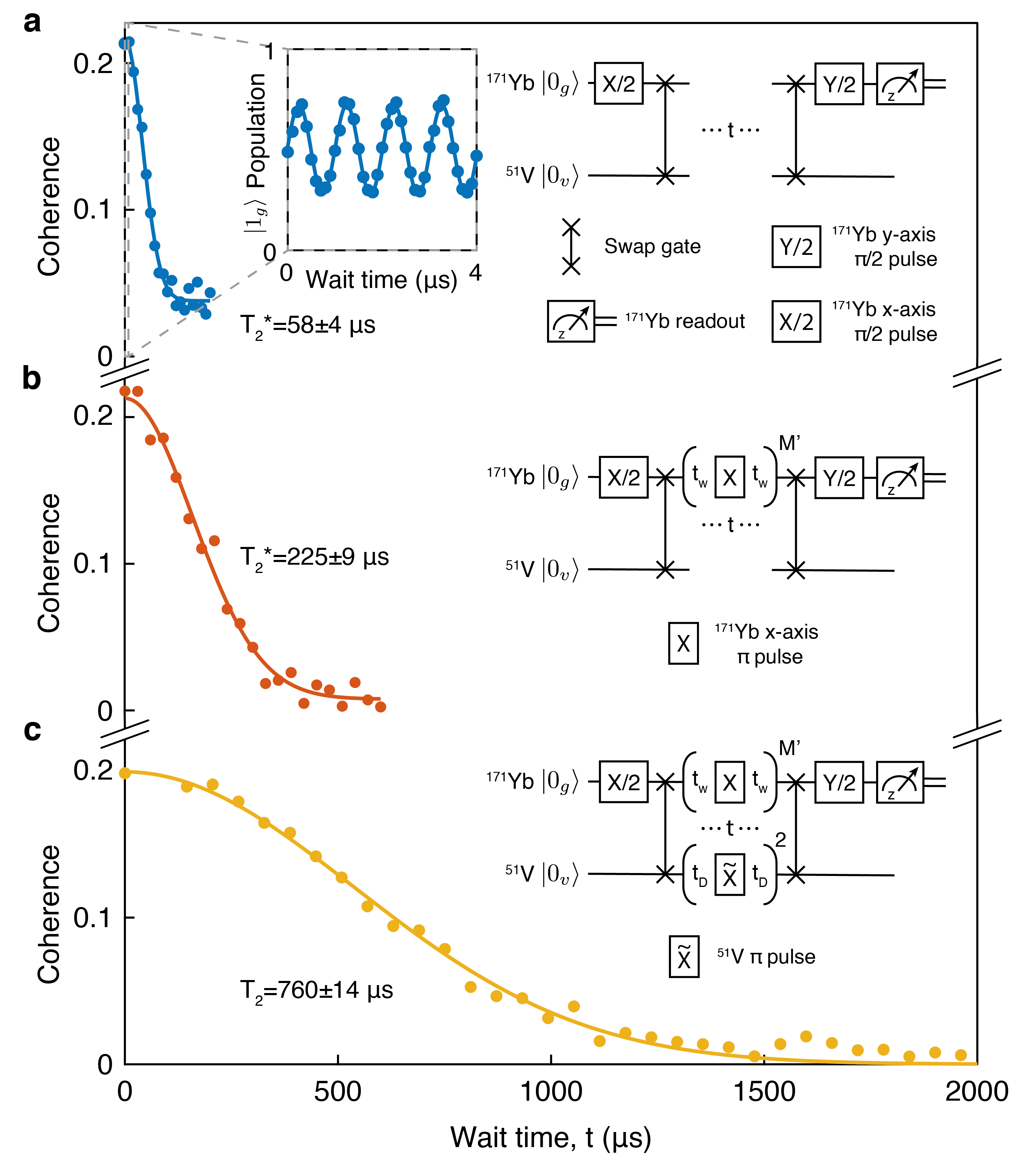}
\caption{{\bf Quantum information storage in the entangled nuclear spin register.} {\bf a,} Ramsey coherence time measurement. The \textsuperscript{171}Yb qubit is prepared in a superposition state which is subsequently swapped onto the \textsuperscript{51}V register. After waiting for a period of time, $t$, the superposition state is swapped back to the \textsuperscript{171}Yb qubit and measured in the $x$ basis. Fast oscillations are observed at the \textsuperscript{51}V $\omega_c/2\pi = 991~$kHz frequency (inset) and the coherence is derived from the oscillation contrast. The resulting $1/e$ coherence decay time is measured to be $58\pm4~\mu$s. Note that the wait time excludes the swap gate duration. {\bf b,} Coherence time extension via motional narrowing of the \textsuperscript{171}Yb Knight field. By applying $x$-axis $\pi$ pulses spaced by $2t_w=6~\mu$s to the \textsuperscript{171}Yb qubit, the coherence time of the \textsuperscript{51}V register is extended to $225\pm9~\mu$s.
{\bf c,} Further coherence enhancement via dynamical decoupling of the \textsuperscript{51}V register. In addition to the $\pi$ pulses acting on \textsuperscript{171}Yb, two $\pi$ pulses are applied to the \textsuperscript{51}V register with a variable inter-pulse delay time, $2t_\text{D}$. This rephases contributions to the detuning from the nuclear Overhauser field and leads to an extended memory time of $760\pm14~\mu$s. Note that even numbers of \textsuperscript{51}V $\pi$ pulses are necessary to return the register to the $\{\ket{0_v},\ket{W_v}\}$ manifold prior to state retrieval. In a-c, solid lines are fits to Gaussian decay.}
\end{figure}
\end{center}

To evaluate the performance of the \textsuperscript{51}V register as a quantum memory, we characterize its information storage times under various conditions. Specifically, we first transfer a superposition state from the \textsuperscript{171}Yb qubit, $\frac{1}{\sqrt{2}}\left(\ket{0_g}+i\ket{1_g}\right)$, to the \textsuperscript{51}V register via the ZenPol-based swap gate. Subsequently, the transferred state $\frac{1}{\sqrt{2}}\left(\ket{0_v}+\ket{W_v}\right)$ is stored for a variable wait time, $t$, before being swapped back to the \textsuperscript{171}Yb and measured along the $x$-axis, thereby probing the coherence of the final state. As shown in Fig.~3a, we observe a sinusoidal oscillation of the \textsuperscript{171}Yb population, modulated by a Gaussian coherence decay, whose contrast vanishes with a $1/e$ time of $T_2^*=58\pm4~\mu$s. This oscillation has a frequency of $\omega_c/2\pi = 991~$kHz, originating from relative phase accumulation between $\ket{0_v}$ and $\ket{W_v}$ during the wait time. The coherence time of the \textsuperscript{51}V register is predominantly limited by local magnetic field noise from two sources: a fluctuating \textsuperscript{171}Yb dipole moment (\textsuperscript{171}Yb Knight field) and the nuclear Overhauser field (Supplementary Information). As shown in Fig.~3b, the noise created by \textsuperscript{171}Yb can be effectively decoupled from the register by periodically flipping the \textsuperscript{171}Yb magnetic dipole orientation via a series of $\pi$ pulses. Similar to the motional narrowing effect \cite{Bauch2018}, the neutralization of the dipole moment arrests undesired phase diffusion of the register, leading to an increased $1/e$ coherence time of $T_2^*=225\pm9~\mu$s. We further extend the coherence time by performing dynamical decoupling on the \textsuperscript{51}V register to mitigate the decoherence effect of the nuclear spin bath. This relies on applying \textsuperscript{51}V $\pi$ pulses resonant with the $\omega_c$ transition whilst leaving the bath unperturbed (Extended~Data~Fig.~7 and Methods). In Fig.~3c, we apply two \textsuperscript{51}V $\pi$ pulses with variable inter-pulse delay, combined with periodic $\pi$ pulses applied to the \textsuperscript{171}Yb qubit, significantly extending the $1/e$ coherence time to $T_2=760\pm14~\mu$s.

We also characterise the population relaxation times of the $\ket{0_v}$ and $\ket{W_v}$ states with measured lifetimes of $T_1^{(0)}=0.54\pm0.08$ s and $T_1^{(W)}=39.5\pm1.3~\mu$s, respectively. Due to the entangled nature of the $\ket{W_v}$ state, $T_1^{(W)}$ is limited by dephasing and is extended to $127\pm8~\mu$s and $640\pm20~\mu$s by applying the same decoupling sequences as in Fig.~3b,c respectively (Extended~Data~Fig.~8). We note that these dephasing processes can be sensitive to the the stochastic occupation of the $\ket{+m_I}$ and $\ket{-m_I}$ states, depending on the degree of noise correlation between the four register spins (Supplementary Information).

\begin{center}
\begin{figure}[t!]
\includegraphics[width=0.48\textwidth]{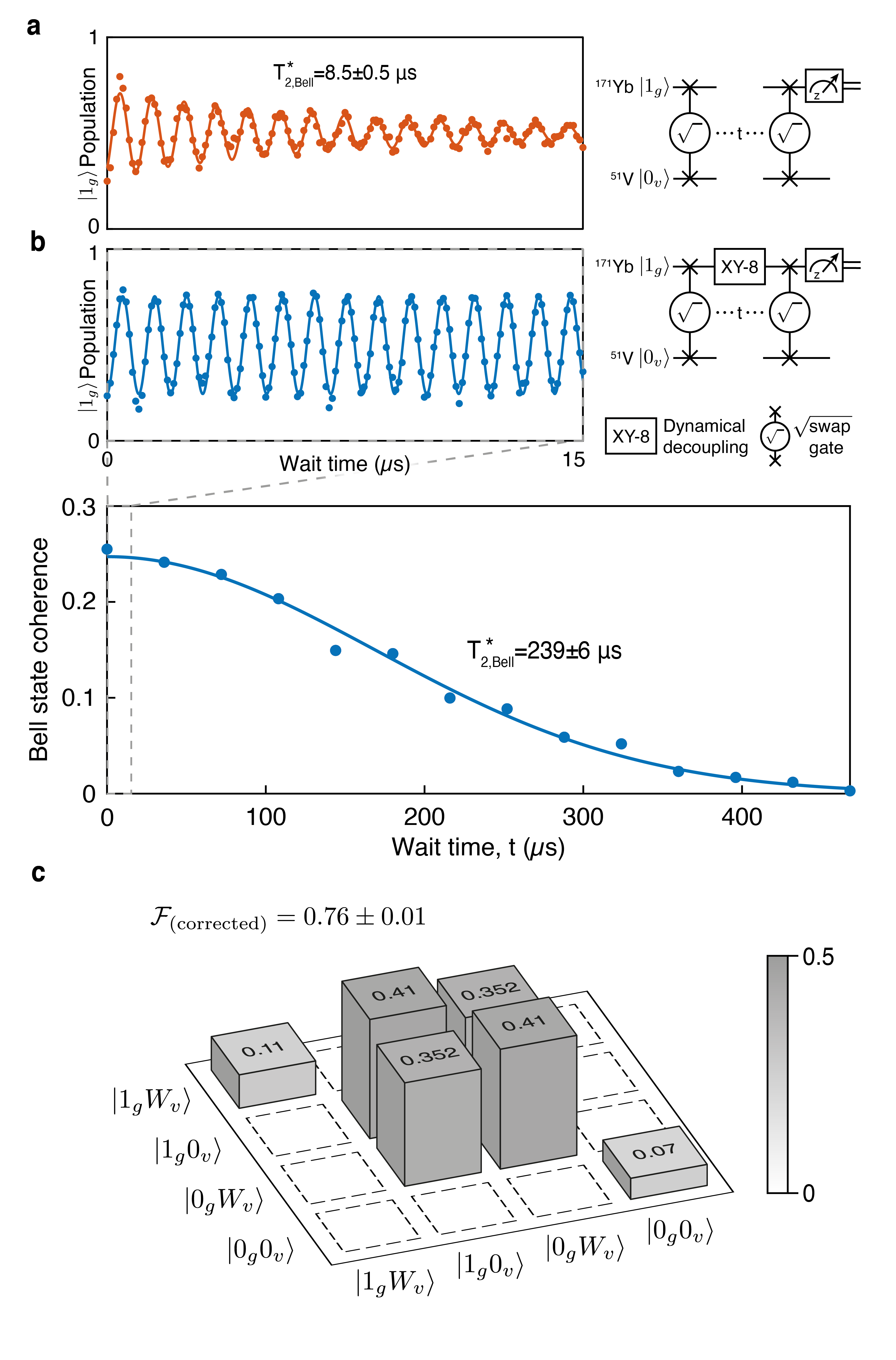}
\caption{{\bf Characterization of maximally entangled \textsuperscript{171}Yb--\textsuperscript{51}V register Bell state.} {\bf a,} Parity oscillations between $\ket{\Psi^+}$ and $\ket{\Psi^-}$ (where $\ket{\Psi^{\pm}} = 1/\sqrt{2}(\ket{1_g}\ket{0_v} \mp i\ket{0_g} \ket{W_v})$) revealing the Bell state coherence time. To prepare the $\ket{\Psi^+}$ Bell state, a $\sqrt{\text{swap}}$ gate is applied to $\ket{1_g}\ket{0_v}$; subsequently during a wait time of duration $t$ coherent parity oscillations occur between $\ket{\Psi^+}$ and $\ket{\Psi^-}$ at the \textsuperscript{51}V $\omega_c$ transition frequency. A second $\sqrt{\text{swap}}$ gate maps the resulting parity to \textsuperscript{171}Yb population. The oscillation contrast (and hence Bell state coherence) decays with a $1/e$ timescale of $T^*_\text{2,Bell}=8.5\pm0.5~\mu$s, consistent with the \textsuperscript{171}Yb $T_2^*$ time. {\bf b,} Bell state coherence extension. During the parity oscillation, we apply an XY-8 decoupling sequence \cite{Gullion1990} to the \textsuperscript{171}Yb qubit. This leads to a significantly extended Bell state coherence time of $T^*_\text{2,Bell}=239\pm6~\mu$s, limited by the \textsuperscript{51}V $T_2^*$ time measured in Fig.~3b. {\bf c,} Reconstructed Bell state density matrix. Diagonal entries representing populations are extracted through a sequential tomography protocol \cite{Kalb2017} (Methods). Off-diagonal matrix elements representing coherences are obtained from the parity oscillation contrast. Note that all density matrix values have been corrected to account for readout error, yielding a fidelity of $0.76\pm0.01$. See Methods for details of the correction procedure.}
\end{figure}
\end{center}

Finally, we benchmark our multi-spin register by characterizing fidelities of \textsuperscript{171}Yb--\textsuperscript{51}V Bell state generation and detection, serving as a vital component of the quantum repeater protocol \cite{Briegel1998}. In particular, the maximally entangled Bell state $\ket{\Psi^+} = \frac{1}{\sqrt{2}}\left(\ket{1_g}\ket{0_v} -i \ket{0_g}\ket{W_v}\right)$ can be prepared by initialising the system in $\ket{1_g}\ket{0_v}$ and applying a $\sqrt{\text{swap}}$ gate based on the ZenPol sequence satisfying $J_\text{ex}t_M = \pi/2$ (equation~\eqref{magnonosc}). The Bell state coherence is evaluated by monitoring the contrast of oscillation between a given Bell state and its parity conjugate \cite{Levine2018}. In our system, the free evolution of $\ket{\Psi^+}$ gives rise to a parity oscillation at frequency $\omega_c$ with $\ket{\Psi^-} = \frac{1}{\sqrt{2}}\left(\ket{1_g}\ket{0_v} +i \ket{0_g}\ket{W_v}\right)$ (Supplementary Information). We read out this oscillation by applying a second $\sqrt{\text{swap}}$ gate to the system, encoding the parity into \textsuperscript{171}Yb population. Figure~4a shows the measured parity oscillations decaying with a $1/e$ time of $T^*_\text{2,Bell}=8.5\pm0.5~\mu$s, limited by the $T_2^*$ dephasing time of the \textsuperscript{171}Yb qubit~\cite{Kindem2020}. To improve the coherence, we apply an XY-8 decoupling sequence \cite{Gullion1990} to the \textsuperscript{171}Yb, leading to an enhanced value of $T^*_\text{2,Bell}=239\pm6~\mu$s (Fig.~4b); this timescale is similar to that in Fig.~3b, indicating that the Bell state coherence is likely limited by the $T_2^*$ dephasing time of the \textsuperscript{51}V register.

In order to estimate the Bell state preparation fidelity, defined as $\mathcal{F} = \langle \Psi^+ | \rho | \Psi^+ \rangle$, we perform a sequential tomography protocol~\cite{Kalb2017} to reconstruct the system density matrix $\rho$ in the effective manifold spanned by four states \{$\ket{0_g0_v}, \ket{0_gW_v}, \ket{1_g0_v}, \ket{1_gW_v}$\} (Extended~Data~Fig.~9 and Methods). Taking into account errors in state readout, we obtain a corrected Bell state fidelity of 0.76$\pm$0.01, as summarized in Fig.~4c (the uncorrected fidelity is measured to be 0.61$\pm$0.01). We speculate that this is limited by a combination of incomplete register initialisation, imperfect Hamiltonian engineering and detrimental dephasing of the register during Bell state generation. See Methods and Supplementary Information for detailed discussions including error analysis.

In this work we have demonstrated a noise-robust control protocol to coherently manipulate the local \textsuperscript{51}V nuclear ensemble surrounding a single optically-addressed \textsuperscript{171}Yb spin, enabling the polarisation of the high spin ($I=7/2$) nuclear register, the creation of collective spin-wave excitations, and the preparation of maximally entangled Bell states. Based on these capabilities, we show that the local nuclear spins realise an ensemble-based quantum memory exhibiting long coherence times. Crucially, this memory is {\it deterministic} and {\it reproducible} in that every \textsuperscript{171}Yb ion doped into a YVO\textsubscript{4} crystal accesses a near-identical nuclear register in its local environment (Extended~Data~Fig.~10). We envisage that this resource will enable the implementation of multi-node quantum network architectures using rare-earth ions with both enhanced connectivity and large-scale entanglement \cite{Briegel1998}. Furthermore, realising coherent quantum systems using dense lattice nuclear spins will open the door to exploration of new materials for quantum information applications \cite{Wolfowicz2021}. Finally, these multi-level nuclear spin ensembles offer an attractive, highly controllable platform to investigate the many-body dynamics of a much larger Hilbert space, paving the way for application of solid-state, noisy intermediate-scale quantum (NISQ) devices in the context of quantum simulation~\cite{Gangloff2020,Randall2021}.

\bibliographystyle{naturemag}
\bibliography{YbV_citations}

\clearpage
\onecolumngrid
\setcounter{figure}{0}    
\renewcommand{\figurename}{Extended Data Fig.}
\section*{Extended Data}
\begin{figure}[h!]
\centering
\includegraphics{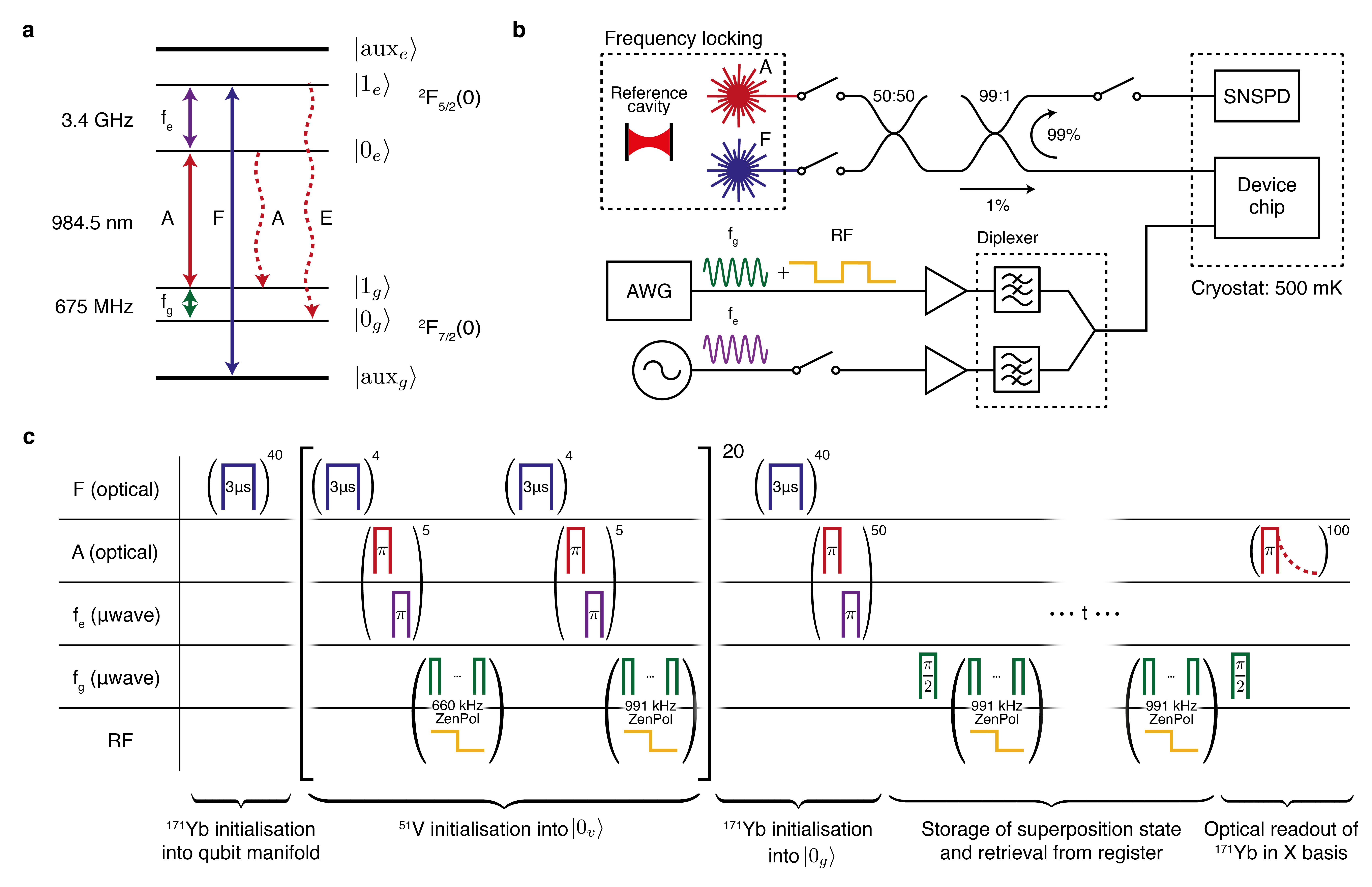}
\caption{{\bf Experimental setup and sequence detail}.
{\bf a,} Energy level structure of \textsuperscript{171}Yb\textsuperscript{3+}:YVO\textsubscript{4} \textsuperscript{2}F\textsubscript{7/2}(0) and \textsuperscript{2}F\textsubscript{5/2}(0). Initialisation into $\ket{0_g}$ involves repeated pulses on the F transition combined with consecutive pairs of $\pi$ pulses applied to the A and f\textsubscript{e} transitions leading to excitation into $\ket{1_e}$. Subsequently, decay via E leads to initialisation into $\ket{0_g}$. Optical readout relies on repeated optical $\pi$ pulses on the A transition, each followed by a photon detection window during which we measure cavity-enhanced emission via A. {\bf b,} Experimental setup. Optical control of the A and F transitions is realised via two frequency-stabilised lasers, each modulated using acousto-optic modulator (AOM) shutters. Microwave control is divided into two paths: a low frequency path consisting of 675 MHz ground state control (f\textsubscript{g} transition) and RF, both generated using a single arbitrary waveform generator (AWG) channel and a high frequency path consisting of 3.4 GHz excited state microwave control (f\textsubscript{e} transition). Each path is independently amplified and combined using a diplexer. The device chip and a superconducting nanowire single photon detector (SNSPD) are cooled to $\approx500$~mK in a cryostat. {\bf c,} Detailed pulse sequence used for quantum state storage and retrieval. First, the $^{51}$V register and $^{171}$Yb qubit are initialised into $\ket{0_v}$ and $\ket{0_g}$, respectively, as described in the main text. Subsequently, the $^{171}$Yb is prepared in a superposition state, via a $\pi/2$ pulse, which is swapped onto the $^{51}$V register using a ZenPol sequence resonant with the 991 kHz $\omega_c$ $^{51}$V transition. After a wait time, $t$, the state is swapped back to $^{171}$Yb and measured in the $x$ basis via a $\pi/2$ pulse followed by optical readout.}
\end{figure}

\begin{figure}[h!]
\centering
\includegraphics{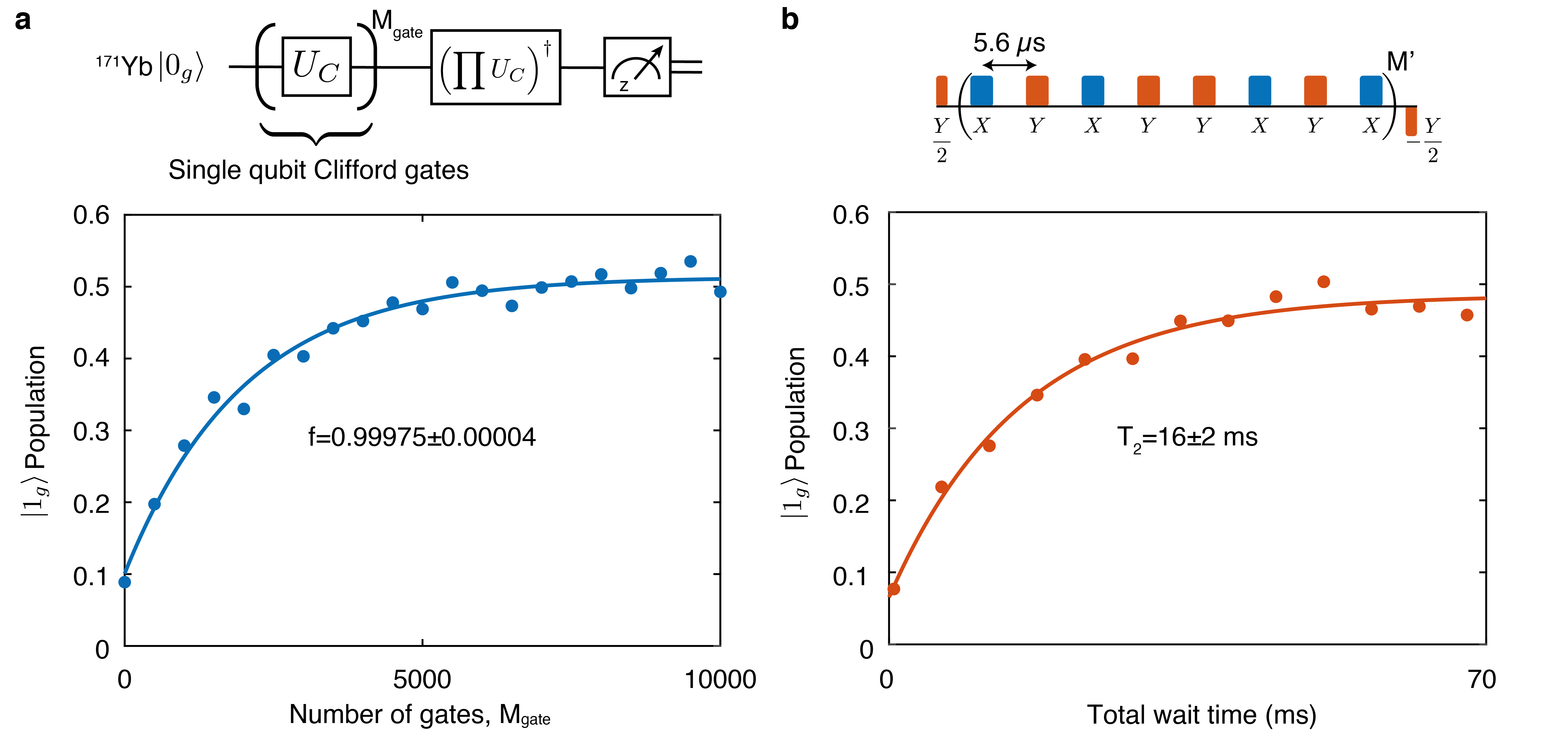}
\caption{{\bf Randomised benchmarking and \textsuperscript{171}Yb qubit coherence}.
\textbf{a,} We measure the average fidelity of single qubit gates applied to the $^{171}$Yb $\ket{0_g}\leftrightarrow\ket{1_g}$ transition. We apply a series of $M_\text{gate}$ randomly sampled Clifford gates followed by the inverse operation (top inset). When averaged over a sufficiently large number of samples (in our case 100) we can extract an average gate fidelity from the $1/e$ exponential decay constant, leading to $f=0.99975\pm0.00004$. \textbf{b,} We also measure the coherence time of the qubit transition using an XY-8 dynamical decoupling pulse sequence (top inset) with a fixed inter-$\pi$ pulse separation of $5.6~\mu$s and variable number of repetitions, $M'$. This leads to an exponential decay with $1/e$ time constant $T_2=16\pm2~$ms.
}
\end{figure}

\begin{figure}[h!]
\centering
\includegraphics{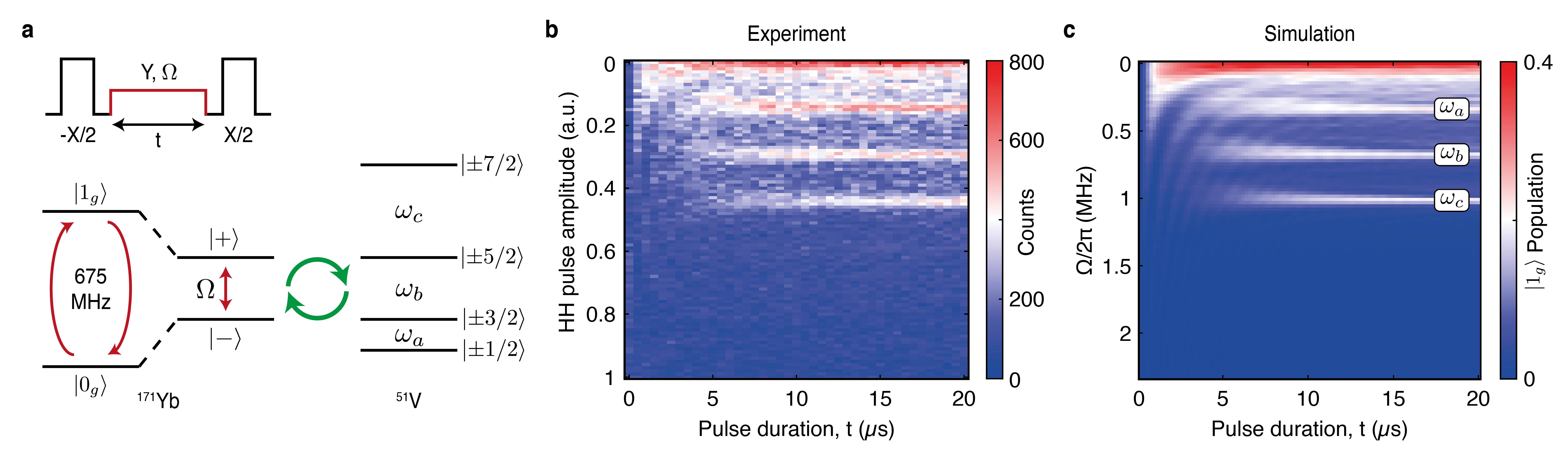}
\caption{{\bf Hartmann Hahn spectroscopy}.
{\bf a,} Hartmann Hahn (HH) sequence used to perform spectroscopy of the nuclear spin environment. During the HH pulse (red), the \textsuperscript{171}Yb $\ket{0_g}\leftrightarrow\ket{1_g}$ qubit transition is driven resonantly for duration $t$ with $y$-phase leading to a pair of dressed states, $\ket{\pm} = \frac{1}{\sqrt{2}}(\ket{0_g} \pm \ket{1_g})$, separated by energy splitting equal to the Rabi frequency, $\Omega$. An initial $-x$-phase $\pi/2$ pulse prepares the $^{171}$Yb qubit in the $\ket{-}$ dressed state. When the Rabi frequency of the HH pulse is tuned to equal one of the \textsuperscript{51}V transition frequencies, the $^{171}$Yb is transferred into the $\ket{+}$ dressed state as a result of resonant population exchange (green arrows). The $\ket{+}$ state population is mapped to $\ket{1_g}$ with a final $x$-phase $\pi/2$ pulse for readout.
{\bf b,}  HH spectroscopy experimental results. To identify nuclear spin resonances, both the HH pulse amplitude and duration are varied. The three evenly-spaced horizontal resonance features occurring at pulse amplitudes of 0.15, 0.3, and 0.45 (in arbitrary units, a.u.) correspond to interaction with the $\omega_a$, $\omega_b$ and $\omega_c$ transitions, respectively. In the no driving ($\Omega = 0$) case, the sequence probes the decoherence dynamics of the prepared $\ket{-}$ state i.e. it measures the Ramsey coherence time.
{\bf c,} HH spectroscopy simulation results. Simulation results agree well with the experiment, corroborating that \textsuperscript{171}Yb--\textsuperscript{51}V interactions are dominant in our system.}
\end{figure}

\begin{figure}[h!]
\centering
\includegraphics{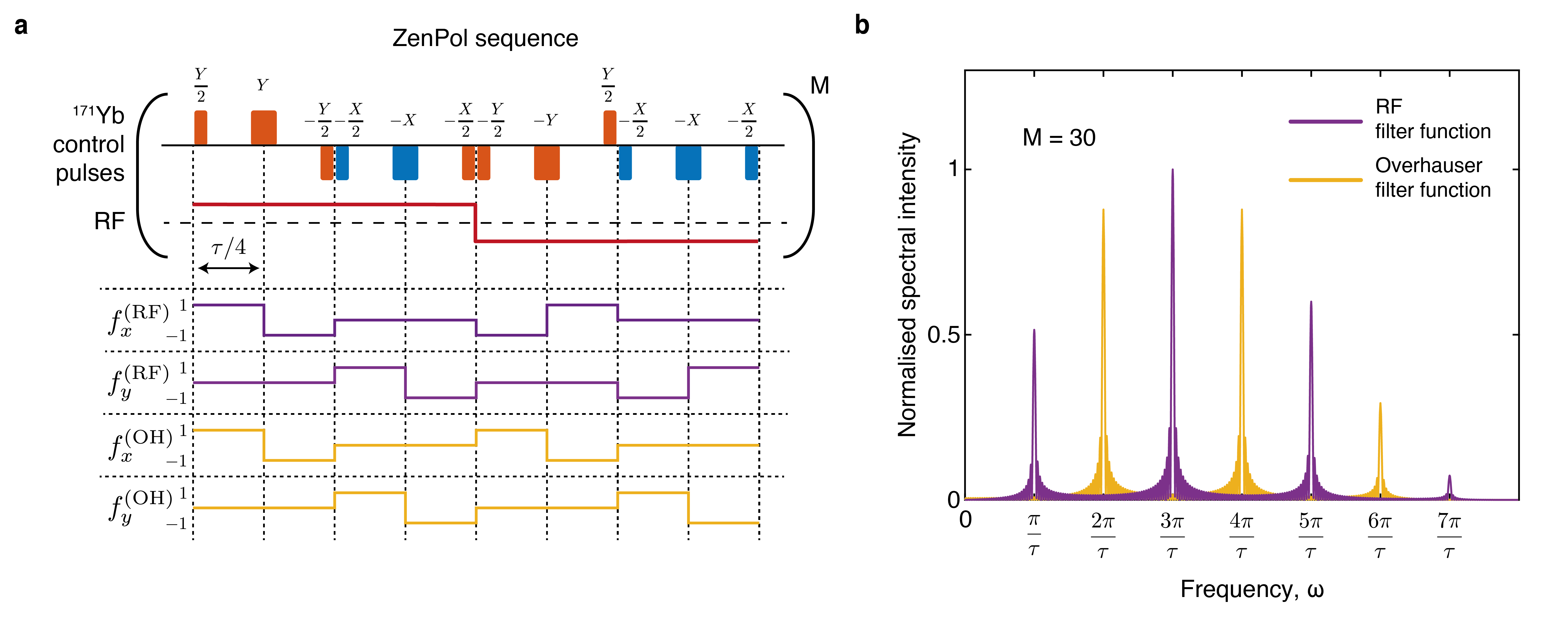}
\caption{{\bf ZenPol sequence detail}.
{\bf a,} ZenPol sequence with the toggling-frame transformation of the $\hat{\tilde{S}}_z$ operator for the $^{171}$Yb qubit. The ZenPol sequence consists of a series of $\pi$ and $\pi/2$ pulses about the $x$- and $y$-axes combined with a synchronously applied, square-wave RF magnetic field with period $2\tau$. The Overhauser- and RF-induced interactions are determined by the toggling-frame transformations of $\Hat{\tilde{S_z}}$ which are given by $\Hat{\tilde{S_x}}f_x^\text{(OH)}+\Hat{\tilde{S_y}}f_y^\text{(OH)}$ and $\Hat{\tilde{S_x}}f_x^\text{(RF)}+\Hat{\tilde{S_y}}f_y^\text{(RF)}$, respectively (see yellow and purple lines for $f_{x,y}^\text{(OH)}$ and $f_{x,y}^\text{(RF)}$, respectively). At the resonance condition $1/2\tau=\omega_j/2\pi k$ for odd integer $k$ with $^{51}$V spin precession frequency $\omega_j$, the sequence realises noise-robust spin-exchange interaction with a time-averaged Hamiltonian that only depends on the RF magnetic field amplitude.
{\bf b,}
ZenPol sequence filter functions corresponding to the Fourier transforms of $f_x^\text{OH}$ (yellow) and $f_x^\text{RF}$ (purple). For a sequence with fixed $\tau$, the peak positions determine the resonant frequencies at which $^{171}$Yb--$^{51}$V interactions can occur. Note that the incoherent Overhauser-induced interactions occur at even-$k$ resonances and are spectrally separated from the coherent RF-induced interactions occurring at odd-$k$ resonances.}

\end{figure}

\begin{figure}[h!]
\centering
\includegraphics{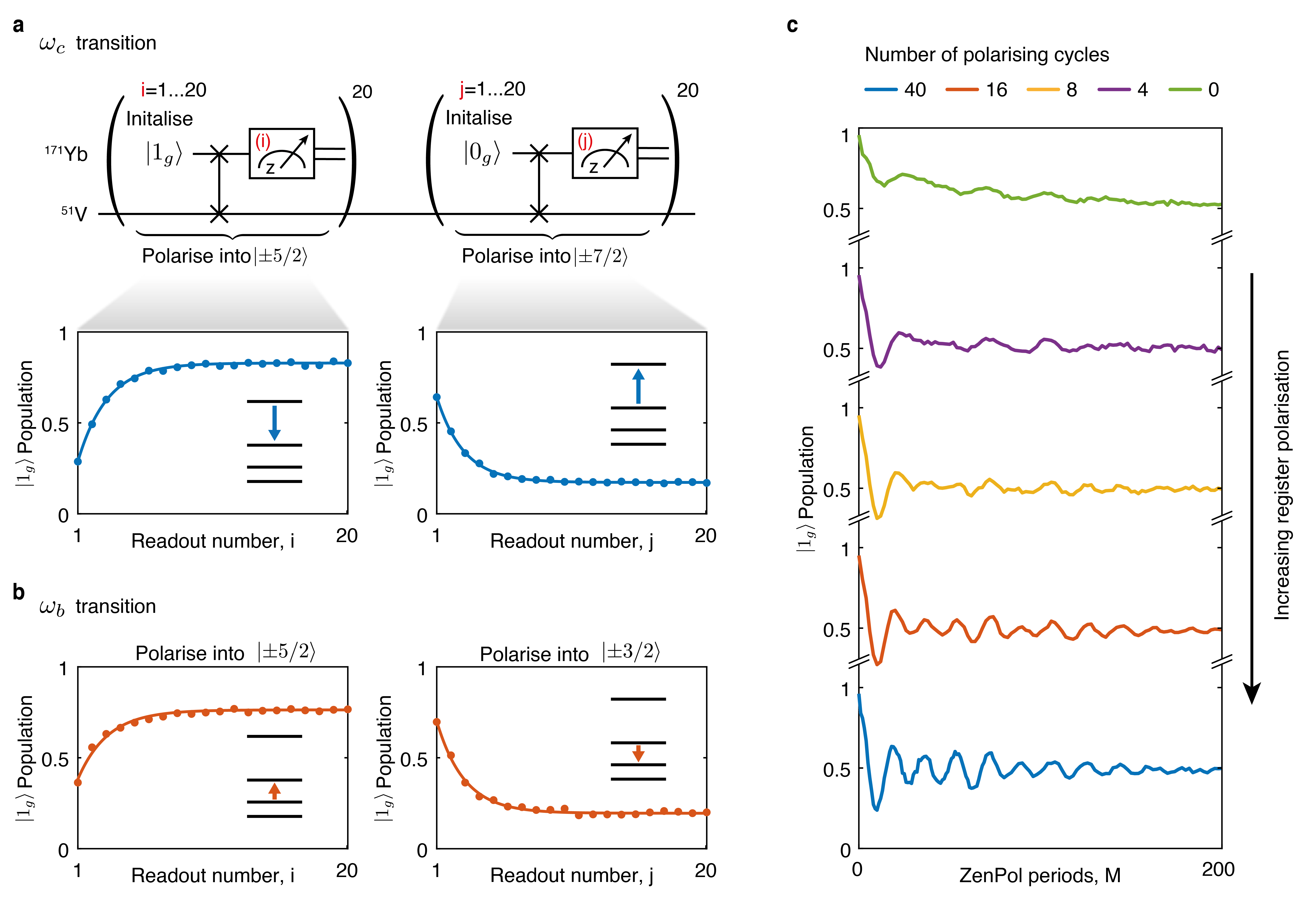}
\caption{{\bf Polarisation of multi-level nuclear register spins}.
{\bf a,} Polarisation readout by polarisation inversion (PROPI) experiments for the $^{51}$V register $\omega_c$ transition. The PROPI sequence performs a repeated swap operation based on the ZenPol sequence, periodically interleaved with $^{171}$Yb qubit readout and re-initialisation into $\ket{1_g}$. A total of 20 polarising cycles are applied to the $\omega_c$ transition to polarise the $^{51}$V register into $\ket{\pm 5/2}$. As a result of register polarisation, the $^{171}$Yb population in $\ket{1_g}$ increases over time, indicating the accumulation of the $^{51}$V population in $\ket{\pm 5/2}$ (left panel). We observe that the register polarisation saturates after approximately 10 cycles. Subsequently, we perform repolarisation cycles where $^{171}$Yb is initialised into $\ket{0_g}$ and $^{51}$V register spins are transferred to $\ket{\pm7/2}$ with similar saturation timescale (right panel). 
{\bf b,} PROPI experiments for the $^{51}$V register $\omega_b$ transition. Applying a ZenPol sequence resonant with the $\omega_b$ transition, interleaved with $^{171}$Yb initialisation into $\ket{1_g}$ ($\ket{0_g}$), results in  $^{51}$V register polarisation into $\ket{\pm5/2}$ ($\ket{\pm3/2}$), as indicated by an increase (decrease) in $^{171}$Yb $\ket{1_g}$ population. 
{\bf c,} Experimental results of ZenPol spin-exchange dynamics with varying degree of $^{51}$V register polarisation. As the number of polarisation cycles used to prepare the $\ket{0_v} = \ket{\pm7/2}^{\otimes 4}$ state increases, the subsequent spin-exchange oscillations become more pronounced. Note that these polarisation cycles are interleaved between the $\omega_b$ and $\omega_c$ transitions.}
\end{figure}

\begin{figure}[h!]
\centering
\includegraphics{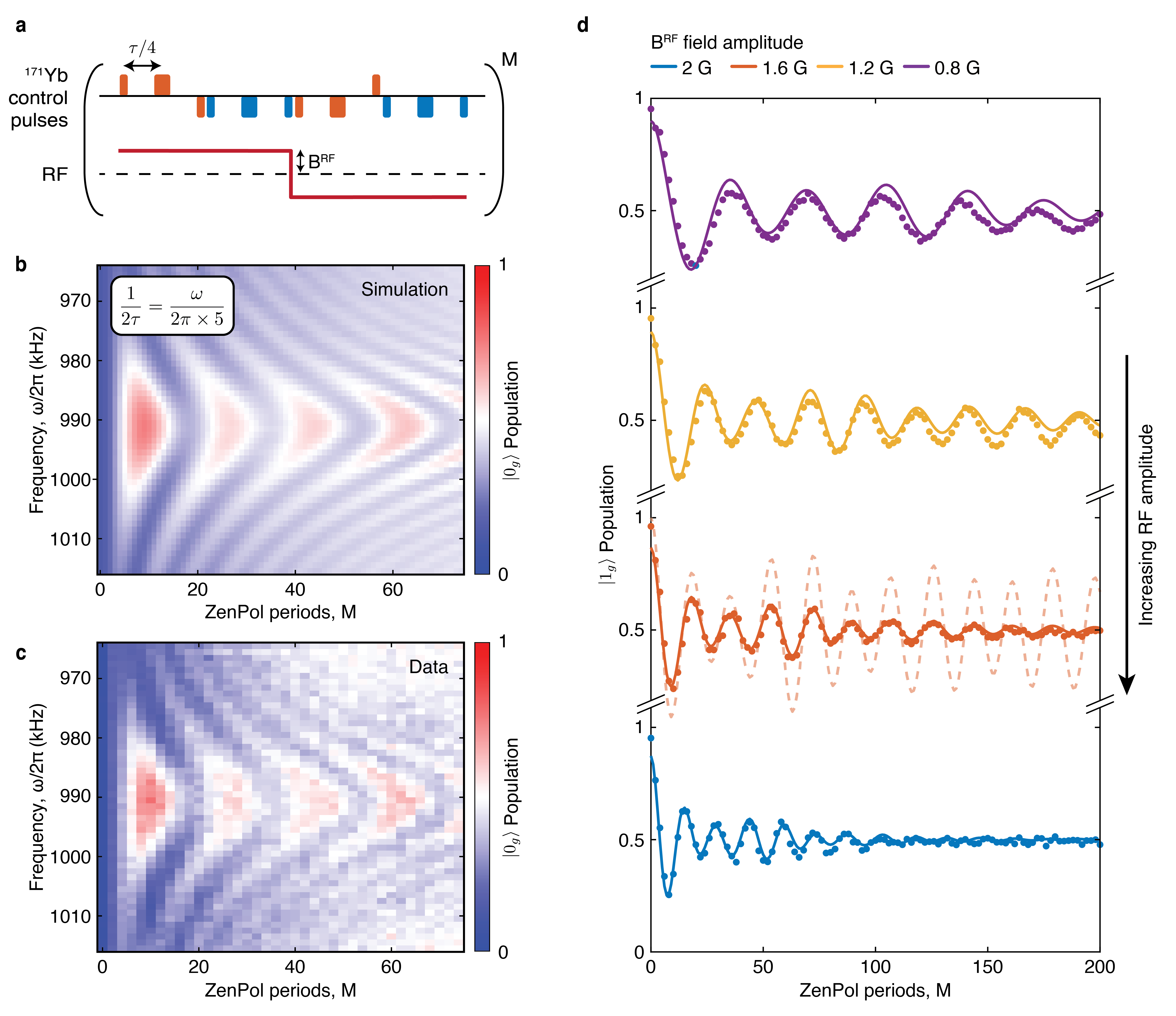}
\caption{{\bf Tunable spin-exchange rate}.
{\bf a,} ZenPol sequence schematic. The square-wave RF magnetic field amplitude $B$\textsuperscript{RF} determines the $^{171}$Yb--$^{51}$V interaction strength, the pulse spacing $\tau/4$ varies the sequence detuning from a specific $^{51}$V nuclear spin transition, and the number of ZenPol periods, $M$, determines the total interaction time.
{\bf b,} Simulated spin-exchange dynamics near the $\omega_c$ transition at $k=5$, probed as a function of sequence resonance frequency $\omega$ and the number of ZenPol periods, $M$.
{\bf c,} Measured spin-exchange dynamics showing good agreement with the numerical simulation in b.
{\bf d,} Experimental demonstration of tunable spin-exchange rate by varying $B^\text{RF}$.
When increasing $B^\text{RF}$ from 0.8~G to 2.0~G, we observe a corresponding linear increase in the spin-exchange rate.
In all cases, numerical simulations (solid lines) taking into account incomplete register polarisation, control pulse imperfections and an exponential phenomenological decay show reasonable agreement with the experimental data (markers). A simulation result without this phenomenological decay (dashed line) displays a discrepancy, which needs further investigation. See Supplementary Information for simulation details.
}
\end{figure}

\begin{figure}[h!]
\centering
\includegraphics{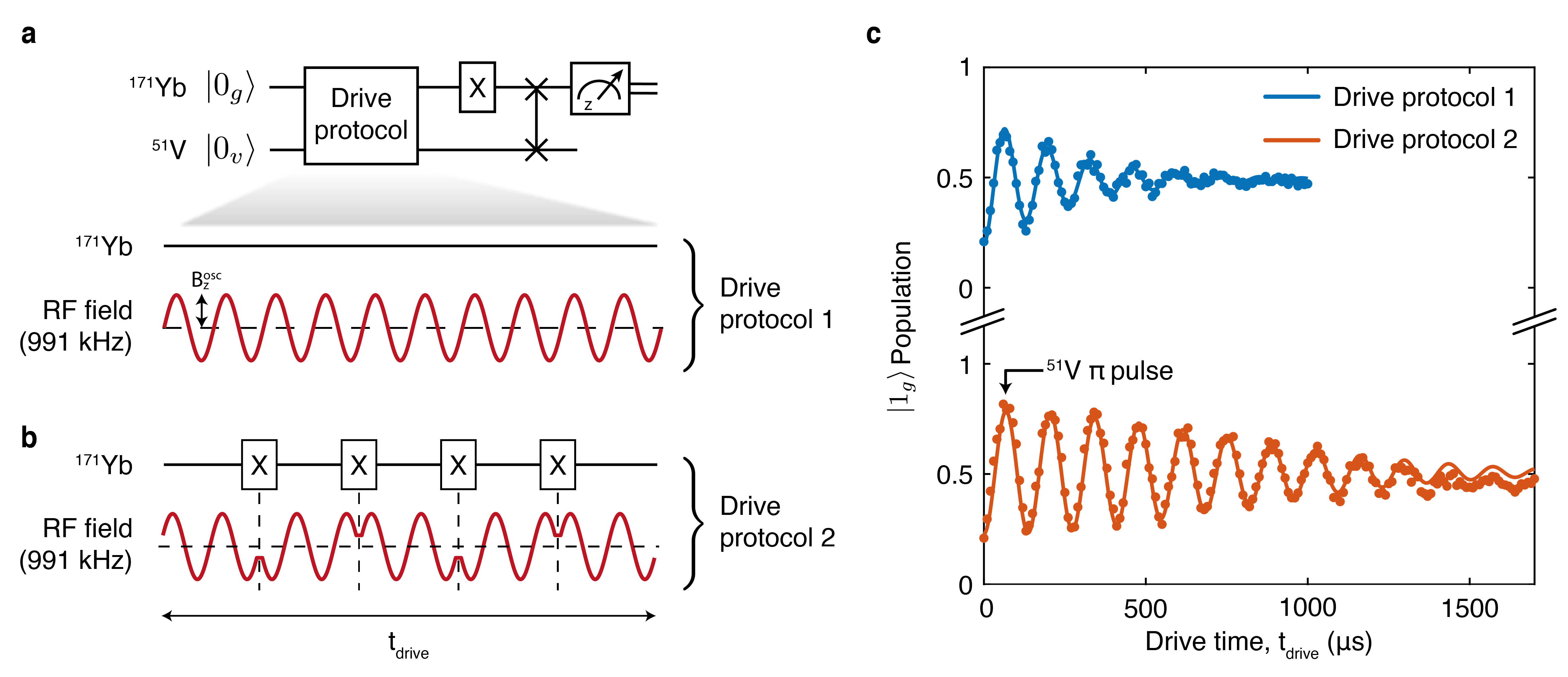}
\caption{{\bf Direct $^{51}V$ nuclear spin driving}.
{\bf a,} Details of \textsuperscript{51}V nuclear spin driving scheme. To directly drive the $^{51}$V nuclear spin $\omega_c$ transition, a sinusoidal $z$-directed RF magnetic field, $B_z^\text{osc}\sin(\omega_c t)$, is applied to the system at a frequency of $\omega_c/2\pi = 991$~kHz after initialising the $^{171}$Yb and $^{51}$V register into $\ket{0_g}$ and $\ket{0_v} = \ket{\downarrow\downarrow\downarrow\downarrow}$, respectively (Drive Protocol 1). This induces an oscillating magnetic dipole moment on the $^{171}$Yb qubit which in turn generates an amplified transverse driving field at each $^{51}$V (Methods). Consequently, the four $^{51}$V register spins undergo independent Rabi oscillation between the $\ket{\uparrow} = \ket{\pm5/2}$ and $\ket{\downarrow} = \ket{\pm7/2}$ states. To probe the nuclear spin Rabi oscillation, the $\ket{\downarrow}$ population is measured by preparing the $^{171}$Yb in $\ket{1_g}$ via an $x$-phase $\pi$ pulse, performing a single swap gate and reading out the $^{171}$Yb population. {\bf b,} Decoupling of magnetic field noise originating from the $^{171}$Yb Knight field. To improve the nuclear spin control fidelity, a train of equidistant $\pi$ pulses are applied to the $^{171}$Yb during the driving period, thereby cancelling dephasing due to the $^{171}$Yb Knight field (Drive Protocol 2). Each $\pi$ pulse is accompanied by a $\pi$ phase shift of the sinusoidal field to ensure phase continuity of the nuclear Rabi driving and an even number of $\pi$ pulses ensures the $^{171}$Yb state is returned to $\ket{0_g}$ at the end of the sequence (Methods).
{\bf c,} Measured $^{51}$V register Rabi oscillations using the aforementioned schemes. We observe coherent nuclear Rabi oscillations between the $\ket{\downarrow}$ and $\ket{\uparrow}$ states at a Rabi frequency of $2\pi\times (7.65\pm0.05)$ kHz. An exponential decay is observed with a $1/e$ time constant of $280\pm30~\mu$s without decoupling (blue). The additional $\pi$ pulses applied to the $^{171}$Yb qubit lead to an enhancement in control fidelity, giving a $1/e$ Gaussian decay time of $1040\pm70~\mu$s (red). The black arrow at $t\approx69~\mu$s indicates the $^{51}$V $\pi$ pulse used in Fig.~3c.}
\end{figure}

\begin{figure}[h!]
\centering
\includegraphics{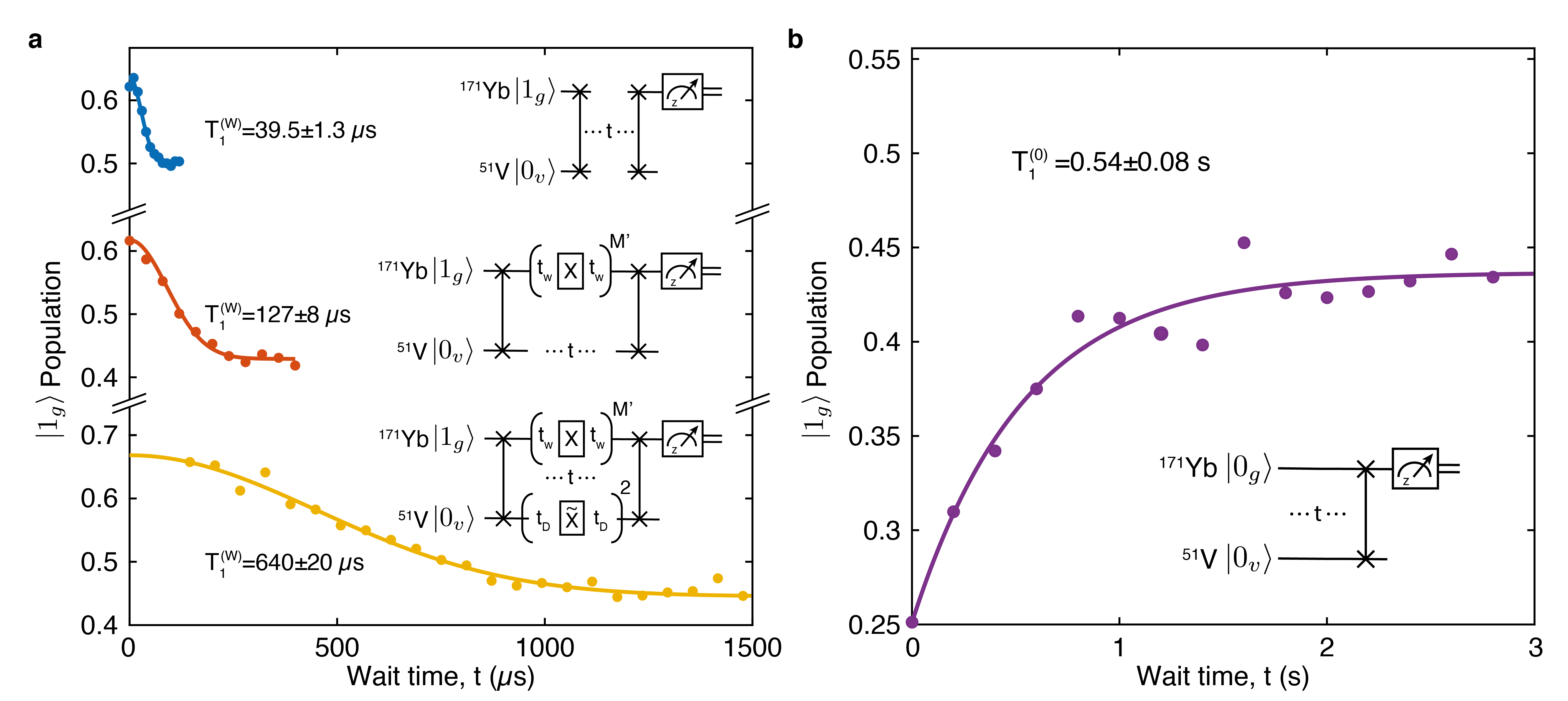}
\caption{{\bf $^{51}$V spin register population relaxation}.
{\bf a,} Measured relaxation timescales,  $T_1^{(W)}$, of the entangled register state, $\ket{W_v}$, under various conditions. Top: the $^{51}$V register is prepared in the $\ket{W_v}$ state by swapping a single spin excitation from the $^{171}$Yb initialised into $\ket{1_g}$. After a variable wait time, $t$, the $^{51}$V state is swapped back onto $^{171}$Yb and measured (top inset). The resulting Gaussian decay shows a $1/e$ relaxation time of  $T_1^{(W)} = 39.5\pm1.3~\mu$s (blue trace), limited by dephasing of the entangled $\ket{W_v}$ state. Middle: the $T_1^{(W)}$ lifetime can be extended by applying a series of equidistant $\pi$ pulses to the $^{171}$Yb separated by $2t_w=6~\mu$s (middle inset). This decouples the $\ket{W_v}$ state from dephasing induced by the $^{171}$Yb Knight field, equivalent to the coherence time extension in Fig.~3b, leading to an extended $1/e$ lifetime of $T_1^{(W)} = 127\pm8~\mu$s (red trace). Bottom: further extension of the $T_1^{(W)}$ lifetime is achieved by dynamical decoupling whereby additionally two $^{51}$V $\pi$ pulses are applied during the wait time with a variable pulse separation $2t_{D}$ (bottom inset). This gives rise to a significantly prolonged lifetime of $T_1^{(W)} = 640\pm20~\mu$s (yellow trace), equivalent to the coherence time extension in Fig.~3c. {\bf b,} Measured relaxation timescale,  $T_1^{(0)}$, of the polarized register state $\ket{0_v}$.The register is initialised in $\ket{0_v}$ and after a variable wait time, $t$, the $^{51}$V state is swapped onto $^{171}$Yb and measured (inset). We observe an exponential decay with a $1/e$ relaxation time of $T_1^{(0)} = 0.54\pm0.08~$s, likely limited by incoherent spin exchange with the bath. See Supplementary Information for detailed discussion of $T_1$ relaxation mechanisms.}
\end{figure}

\begin{figure}[h!]
\centering
\includegraphics[width=1.0\textwidth]{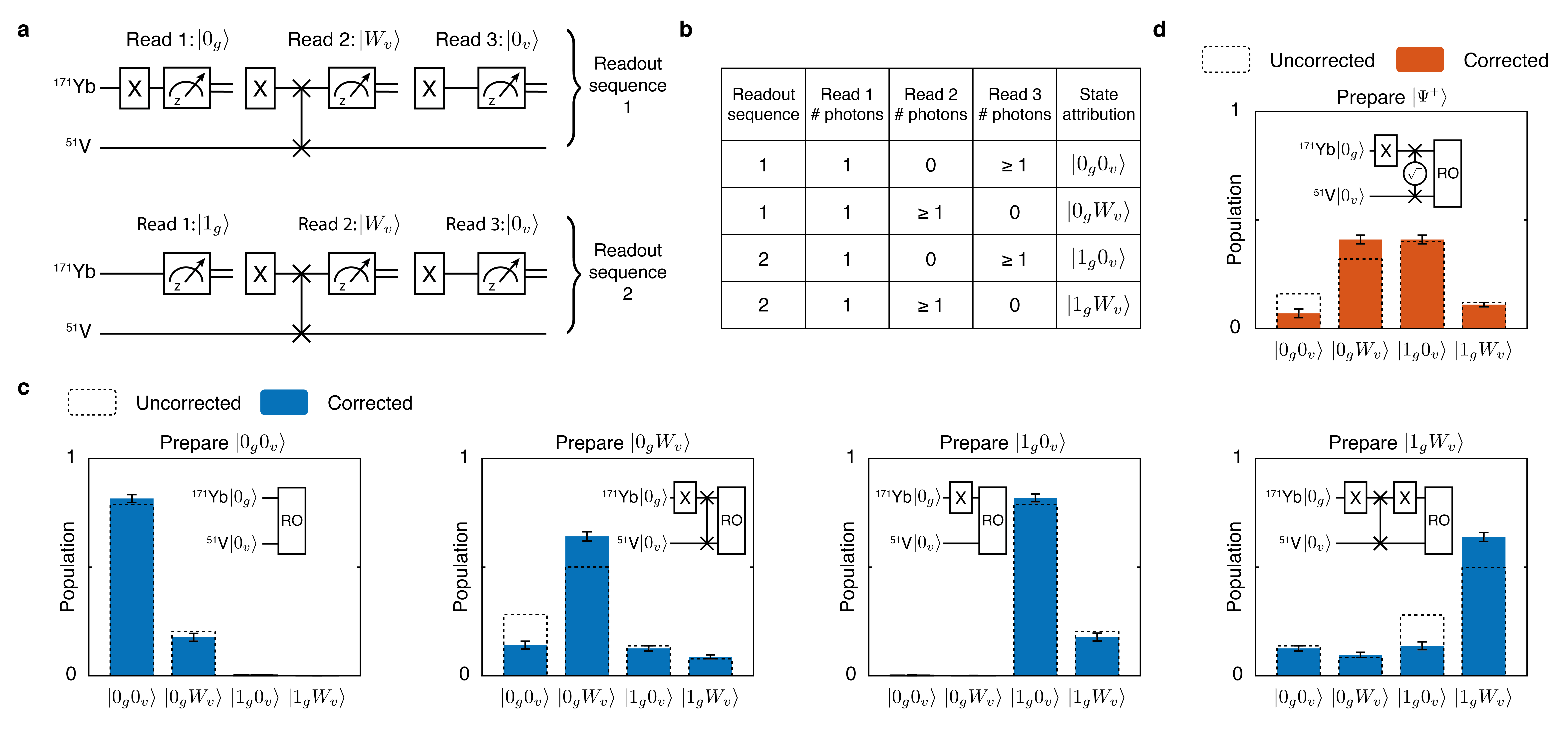}
\caption{{\bf Population measurement histograms for register fidelity characterization}.
{\bf a,} Sequential tomography protocol for characterising $^{171}$Yb--$^{51}$V populations in the basis spanned by \{$\ket{0_g 0_v}$, $\ket{0_gW_v}$, $\ket{1_g0_v}$, $\ket{1_gW_v}$\}. Reconstructing the population probability distribution utilises Readout sequences 1 and 2, each including three consecutive $^{171}$Yb state readouts interleaved with single-qubit gate operations and a swap gate. {\bf b,} Table summarising the post-processing criteria for state attribution. Readout sequences 1 and 2 measure the \{$\ket{0_g0_v}$,$\ket{0_gW_v}$\} and \{$\ket{1_g0_v}$,$\ket{1_gW_v}$\} populations, respectively, conditioned on the three measurement outcomes. See Methods for full details of the post-processing procedure.
{\bf c,} Reconstructed population distributions for estimating state preparation fidelity. The four basis states, \{$\ket{0_g 0_v}$, $\ket{0_gW_v}$, $\ket{1_g0_v}$, $\ket{1_gW_v}$\}, are independently prepared by applying a combination of $^{171}$Yb $\pi$ pulses and swap gates to the initial $\ket{0_g0_v}$ state (see the insets of each subplot). Subsequently, the sequential tomography protocol for state readout (RO) is applied iteratively, alternating between Readout 1 and 2 sequences to fully reconstruct the population probability distributions. {\bf d,} Reconstructed population distribution for the $^{171}$Yb--$^{51}$V Bell state (reproduced from Fig.~4c). The maximally entangled Bell state $\ket{\Psi^+}=1/\sqrt{2}(\ket{1_g0_v}-i\ket{0_gW_v})$ is prepared by applying a $\sqrt{\text{swap}}$ gate to $\ket{1_g0_v}$ and measured using RO (inset). In c,d, the uncorrected and readout-corrected measurement results are presented as dashed and solid filled histograms, respectively. Populations are corrected by accounting for the swap gate error during the readout sequences (Methods).}
\end{figure}

\begin{figure}[h!]
\centering
\includegraphics{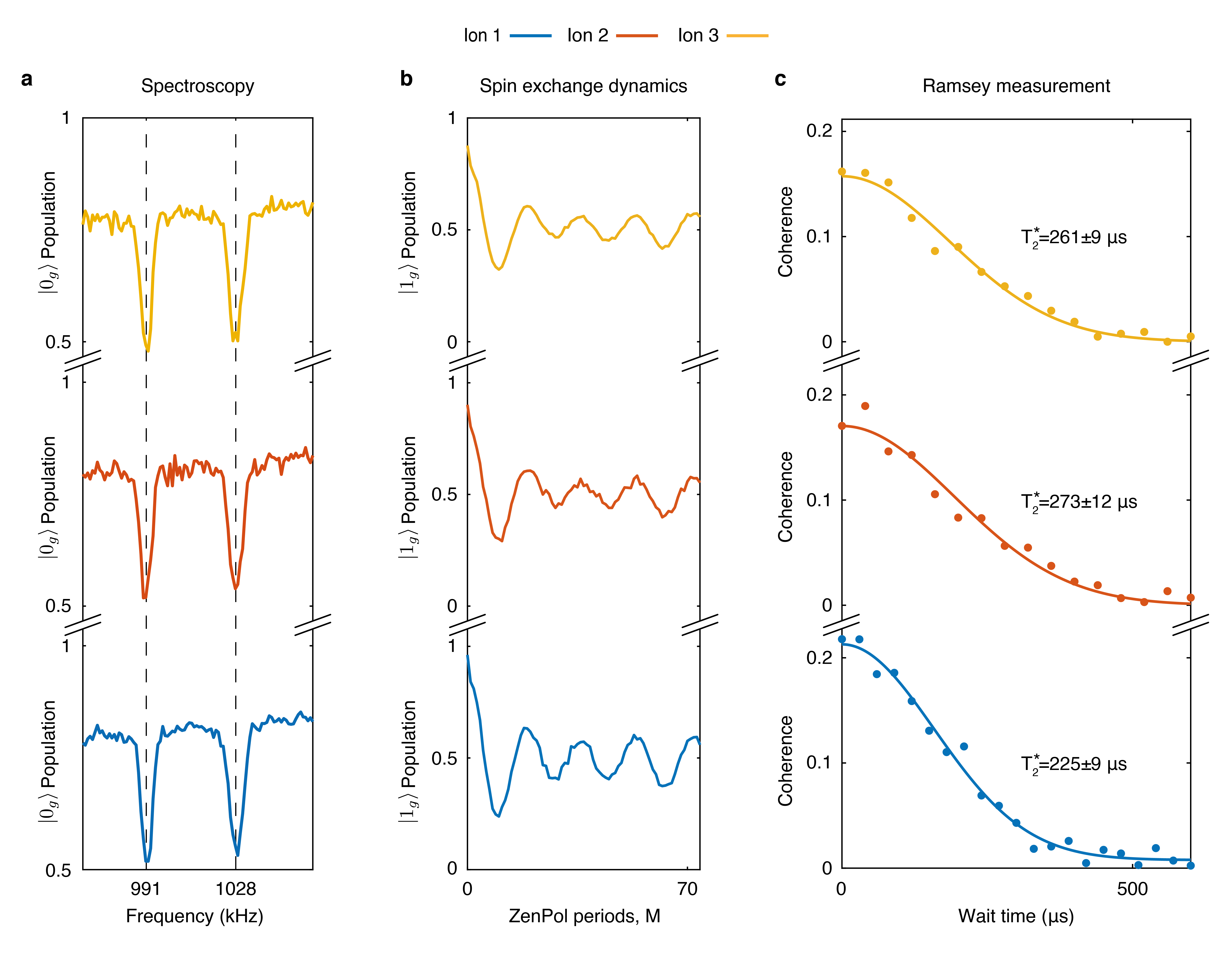}
\caption{{\bf Experimental demonstration of deterministic nuclear spin register}.
To demonstrate the deterministic nature of the nuclear spin register, we perform the same measurements on two additional $^{171}$Yb ion qubits present in the device: Ion 2 (red) and Ion 3 (yellow). Results for Ion 1 (blue) are reproduced from the main text figures for ease of comparison. {\bf a,} ZenPol spectra near the $\omega_c~(k=5)$ resonance of the $^{51}$V register spins. Notice that for all three ions, the bath and register transitions are identified at the same resonance frequencies of $\omega_c^\text{bath}/2\pi = 1028$~kHz and  $\omega_c/2\pi = 991$~kHz, respectively. {\bf b,} Dynamically engineered spin-exchange dynamics between the $^{171}$Yb qubit and $^{51}$V register. Using constant ZenPol square-wave RF amplitude we obtain equal spin-exchange rates for all three ions. {\bf c,} Characterisation of $^{51}$V register coherence times with decoupling from the $^{171}$Yb Knight field. The $1/e$ coherence times are measured to be $225\pm9~\mu$s, $273\pm12~\mu$s and $261\pm9~\mu$s for Ions 1, 2 and 3, respectively. All of these results demonstrate that our platform provides a nearly identical nuclear spin register for every $^{171}$Yb qubit in the system.}
\end{figure}

\clearpage
\twocolumngrid
\section*{Methods}
\subsection*{Experimental Setup}
The YVO\textsubscript{4} crystal used in this project was cut and polished from an undoped boule (Gamdan Optics) with a residual total \textsuperscript{171}Yb concentration of 140 ppb. Nanophotonic cavities were fabricated from this material using focused ion beam milling, see \cite{Zhong2016,Zhong2017a} for more detail on this process. The cavity used in this work has a Q-factor of $\approx10,000$ leading to Purcell enhancement and consequent reduction of the $^{171}$Yb excited state lifetime from 267 $\mu$s to 2.3 $\mu$s as described and measured in \cite{Kindem2020} and $>99\%$ of ion emission coupling to the cavity mode. The reduced optical lifetime enables detection of single $^{171}$Yb ions. The cavity is undercoupled with $\kappa_\text{in}/\kappa\approx0.14$ leading to $14\%$ of emitted light entering the waveguide mode. Waveguide--free-space coupling is achieved via angled couplers with an efficiency of $\approx25\%$ and the end-to-end system efficiency (probability of detecting an emitted photon) is $\approx 1\%$.

The device sits on the still-plate of a $^3$He cryostat (Bluefors LD-He250) with base temperature of 460 mK. Optical signals are fed into the fridge through optical fibre and focused onto the device with an aspheric lens doublet mounted on a stack of $x$-$y$-$z$ piezo nano-positioners (Attocube). The device is tuned on-resonance with the $^{171}$Yb optical transitions via nitrogen condensation. Residual magnetic fields are cancelled along the crystal $c\equiv z$ axis with a set of home-built superconducting magnet coils.

The various optical transitions of a single \textsuperscript{171}Yb qubit are employed for state readout and initialisation (Extended~Data~Fig.~1a). Optical addressing of the A transition for readout is established with a continuous-wave (CW) titanium sapphire (Ti:Sapph) laser (M\textsuperscript{2} Solstis) which is frequency-stabilised to a high-finesse reference cavity (Stable Laser Systems) using Pound-Drever-Hall locking \cite{Drever1983}. The laser double-passes through two free-space acousto-optic-modulator (AOM) setups leading to single-photon level extinction of the input beam, and pulse generation with $\approx10~$ns rise times. A second CW external cavity diode laser (Toptica DL-Pro) is used to address the F transition during initialisation. The laser passes through an identical AOM setup and is frequency stabilised via offset-frequency locking to the Ti:Sapph.

The light output from the cavity is separated from the input with a 99:1 fibre beamsplitter, and passed through a single AOM which provides time-resolved gating of the light to prevent reflected laser pulses from saturating the detector. The light is then sent to a tungsten-silicide superconducting nanowire single photon detector (SNSPD) (Photonspot) which also sits on the still-plate of the cryostat. Photon detection events are subsequently time-tagged and histogrammed (Swabian Timetagger 20).

Microwave pulses to control the ground-state qubit transition (675 MHz) and square-wave RF to generate the $^{171}$Yb--$^{51}$V interaction (100--300 kHz) are directly synthesised with an arbitrary waveform generator (Tektronix 5204AWG) and amplified (Amplifier Research 10U1000). A second microwave path is used for the excited state microwave control (3.4 GHz) necessary for qubit initialisation. The control pulses are generated by switching the output of a signal generator (SRS SG386) and amplifying (Minicircuits ZHL-16W-43-S+). The two microwave signal paths are combined with a diplexer (Marki DPXN2) and sent into the fridge to the device. A gold coplanar waveguide fabricated on the YVO\textsubscript{4} surface enables microwave driving of the ions.

See Extended~Data~Fig.~1 for a schematic of the complete experimental setup.

\subsection*{$^{171}$Yb Initialisation, Readout and Experiment Sequence}
At the 500 mK experiment operating temperature and at zero magnetic field, the equilibrium $^{171}$Yb population is distributed between the $\ket{\text{aux}_g}$, $\ket{0_g}$ and $\ket{1_g}$ states (Extended~Data~Fig.~1a). All experiments start by initialising the single $^{171}$Yb ion into $\ket{0_g}$ via a two-stage protocol~\cite{Kindem2020}. Firstly the $\ket{\text{aux}_g}$ state is emptied with a series of $3~\mu$s pulses applied to the optical F transition each followed by a $3~\mu$s wait period. When the $^{171}$Yb ion is successfully excited from $\ket{\text{aux}_g}$ to $\ket{1_e}$, the population in $\ket{1_e}$ will preferentially decay to $\ket{0_g}$ during the wait time via the cavity-enhanced E transition. Subsequently, the $\ket{1_g}$ state is also emptied by applying an optical $\pi$ pulse to the A transition followed by a microwave $\pi$ pulse to the $f_e$ transition in rapid succession, which similarly leads to excitation from $\ket{1_g}$ to $\ket{1_e}$ and decay into $\ket{0_g}$. This process is repeated several times for improved fidelity.

Readout of the $^{171}$Yb $\ket{1_g}$ state is performed by applying a series of 100 $\pi$ pulses to the A transition, each of which is followed by a $10~\mu s$ photon detection window. This process is enabled by the cyclic nature of the A transition. To read out the $\ket{0_g}$ population we apply an additional $\pi$ pulse to swap the $\ket{0_g} \leftrightarrow \ket{1_g}$ populations before performing the same optical readout procedure.

Extended~Data~Fig.~1c shows an exemplary pulse sequence used to store and retrieve a superposition state from the register consisting of four $^{51}$V lattice ions. The sequence starts with initialisation of the $^{171}$Yb qubit into $\ket{0_g}$ and the spin-7/2 $^{51}$V register into $\ket{0_v} = \ket{\pm7/2}^{\otimes 4}$. A series of ZenPol polarisation operations are interleaved with $^{171}$Yb re-initialisation sequences and alternate between $\omega_b$ and $\omega_c$ transition control to sequentially polarize the $^{51}$V register towards the $\ket{\pm7/2}$ level. After the initialization sequence, a single $\pi/2$ pulse is applied to the $^{171}$Yb qubit to prepare a superposition state. Subsequently, the state is transferred to the $^{51}$V register using a swap operation resonant with the $\omega_c$ transition as detailed in the main text. After a variable wait time, the superposition state is retrieved with a second swap gate and measured in the $x$-basis via a $\pi/2$ pulse followed by optical readout on the A transition as detailed above.

\subsection*{ZenPol Sequence}
We consider a system of a single $^{171}$Yb qubit coupled to four neighbouring nuclear spin-7/2 $^{51}$V ions. This hybrid spin system is described by the effective Hamiltonian (setting $\hbar = 1$):
\begin{multline}
\label{oghamiltonian}
	\hat{H} = \Delta(t) \hat{\tilde{S}}_z + \sum_{i \in \text{register}} Q (\hat{I}_z^{(i)})^2 + \\
	\sum_{i \in \text{register}} \hat{\tilde{S}}_z  \left[B^{\text{OH}}_z+B^\text{RF}(t)\right]\left[ a_{x} \hat{I}^{(i)}_x + a_{z} \hat{I}^{(i)}_z \right] 
\end{multline}
where $\Delta(t) = \gamma_z^2 (B^\text{OH}_{z} + B^\text{RF}(t))^2 / 2\omega_{01}$ is the effective energy shift due to both $z$-directed nuclear Overhauser ($B^\text{OH}_z$) and external RF ($B^\text{RF}(t)$) magnetic fields, $\omega_{01}/2\pi = 675$~MHz is the $^{171}$Yb qubit transition frequency, $\gamma_z/2\pi = 8.5$~MHz/G is the $^{171}$Yb ground-state longitudinal gyromagnetic ratio, $Q/2\pi = 165$~kHz is the $^{51}$V register nuclear quadrupole splitting, $\hat{\tilde{S}}_z$ is the $^{171}$Yb qubit operator along the $z$-axis, $\hat{I}_{x,z}$ are the $^{51}$V spin-7/2 operators along the $x$- and $z$-axis, and $a_{x,z}$ are the effective coupling strengths between $^{171}$Yb and $^{51}$V along the $x$- and $z$-axes. See Supplementary Information for a detailed derivation of this effective Hamiltonian.

As discussed in the main text, polarisation of the $^{51}$V register and preparation of collective spin-wave states relies on induced polarisation transfer from the $^{171}$Yb to $^{51}$V and is achieved via periodic driving of the $^{171}$Yb qubit. Specifically, periodic pulsed control can dynamically engineer the original Hamiltonian (equation~\eqref{oghamiltonian}) to realize effective spin-exchange interaction between $^{171}$Yb and $^{51}$V ions of the form, $\hat{\tilde{S}}_+ \hat{I}_- + \hat{\tilde{S}}_-\hat{I}_+$, in the average Hamiltonian picture \cite{Choi2019},\cite{Slichter1992}. One example of such a protocol is the recently developed PulsePol sequence \cite{Schwartz2018}, however, it relies on states with a constant, non-zero magnetic dipole moment and therefore cannot be used in our system since the $^{171}$Yb qubit has no intrinsic magnetic dipole moment. Motivated by this approach, we have developed a variant of the PulsePol sequence that accompanies a square-wave RF field synchronized with the sequence (Extended~Data~Fig.~4a).  The base sequence has a total of 8 free-evolution intervals with equal duration ($\tau/4$) defined by periodically spaced short pulses and is repeatedly applied to $^{171}$Yb. Following the sequence design framework presented in Ref.~\cite{Choi2019}, we judiciously choose the phase and ordering of the constituent $\pi/2$ and $\pi$ pulses such that the resulting effective interaction has spin-exchange form with strength proportional to the RF magnetic field amplitude ($B^\text{RF}$), whilst decoupling from interactions induced by the Overhauser field ($B_z^\text{OH}$). We also design the sequence to cancel detuning induced by both of these fields and to retain robustness against pulse rotation errors to leading order. We term this new sequence `ZenPol' for `\textbf{ze}ro first-order Zeeman \textbf{n}uclear-spin \textbf{pol}arisation'.

To understand how the ZenPol sequence works, one can consider a toggling-frame transformation of the $^{171}$Yb spin operator along the quantisation axis ($\hat{\tilde{S}}_{z,\text{tog}}(t)$): we keep track of how this operator is transformed after each preceding pulse. For example, the first $\pi/2$ pulse around the $y$-axis transforms $\hat{\tilde{S}}_z$ into $-\hat{\tilde{S}}_x$ and the subsequent $\pi$ pulse around the $y$-axis transforms $-\hat{\tilde{S}}_x$ into $+\hat{\tilde{S}}_x$. Over one sequence period, the toggling-frame transformation generates a {\it time-dependent} Hamiltonian $\hat{H}_\text{tog}(t)$ that is piecewise constant for each of 8 free-evolution intervals, which can be expressed as
\begin{multline}
\hat{H}_\text{tog}(t) =\\
\Delta(t)  \left[ f_x^{\text{OH}}(t) \hat{\tilde{S}}_x + f_y^{\text{OH}}(t) \hat{\tilde{S}}_y \right] + \sum_{i \in \text{register}} Q (\hat{I}_z^{(i)})^2 +\\
						\sum_{i \in \text{register}} B^\text{OH}_z\left[ f_x^{\text{OH}}(t) \hat{\tilde{S}}_x + f_y^{\text{OH}}(t) \hat{\tilde{S}}_y \right]  \left[ a_{x}  \hat{I}^{(i)}_x+ a_{z} \hat{I}^{(i)}_z \right]+\\ 
						\sum_{i \in \text{register}} B^\text{RF}\left[ f_x^{\text{RF}}(t) \hat{\tilde{S}}_x + f_y^{\text{RF}}(t) \hat{\tilde{S}}_y \right]  \left[ a_{x}  \hat{I}^{(i)}_x + a_{z} \hat{I}^{(i)}_z \right].\label{eq:H_tog}
\end{multline}
Here, $f^\text{OH}_{x,y}(t)$ describes the time-dependent modulation of the $^{171}$Yb qubit operator along the $z$ axis ($\hat{\tilde{S}}_{z,\text{tog}}(t)=f^\text{OH}_{x}(t) \hat{\tilde{S}}_x+f^\text{OH}_{y}(t) \hat{\tilde{S}}_y$) (Extended~Data~Fig.~4a). Note that $f^\text{OH}_z(t) = 0$ for all intervals. Since the externally-applied square-wave RF field is constant for each half-sequence period, we can replace $B^\text{RF}(t)$ with the amplitude $B^\text{RF}$ and transfer the time dependence to $f_{x,y}^{\text{OH}}$ by applying sign flips, thus leading to redefined modulation functions $f_{x,y}^{\text{RF}}$ (Extended~Data~Fig.~4a).

The spin-7/2 $^{51}$V ion exhibits three distinct transitions at frequencies $\omega_{a,b,c}$ (Fig.~1b). In the following, we consider an effective spin-1/2 system for the $^{51}$V ions using the $\omega_c$ manifold, $\{\ket{\uparrow} = \ket{\pm5/2}$, $\ket{\downarrow} = \ket{\pm7/2}\}$, with
    $\hat{\tilde{I_x}}=\frac{1}{2}\left(\ket{\uparrow}\bra{\downarrow}+\ket{\downarrow}\bra{\uparrow}\right),
    \hat{\tilde{I_y}}=\frac{1}{2i}\left(\ket{\uparrow}\bra{\downarrow}-\ket{\downarrow}\bra{\uparrow}\right)$ and $\hat{\tilde{I_z}}=\frac{1}{2}\left(\ket{\uparrow}\bra{\uparrow}-\ket{\downarrow}\bra{\downarrow}\right)$. 
In a rotating frame with respect to the target frequency $\omega_c$, the nuclear spin operators become $\hat{\tilde{I}}_x \rightarrow \hat{\tilde{I}}_x \cos(\omega_c t) + \hat{\tilde{I}}_y \sin(\omega_c t)$ and $\hat{\tilde{I}}_z \rightarrow \hat{\tilde{I}}_z$. Thus, the leading-order average Hamiltonian, $\hat{H}_\text{avg} = \frac{1}{2\tau} \int_0^{2\tau} dt \; \hat{H}_\text{tog}(t) $, in the rotating frame is given by:
\begin{multline}
	\hat{H}_\text{avg} =  \sum_{i \in \text{register}} \frac{a_x\sqrt{7}}{2\tau} \int_0^{2\tau} dt\bigg\{ \\ B^\text{OH}_z\!\left[ f_x^{\text{OH}}(t) \hat{\tilde{S}}_x {+} f_y^{\text{OH}}(t) \hat{\tilde{S}}_y \right]\!\left[ \hat{\tilde{I}}^{(i)}_x \cos(\omega_c t) {+} \hat{\tilde{I}}^{(i)}_y \sin(\omega_c t) \right] {+}\\
						B^{\text{RF}}\!\left[ f_x^{\text{RF}}(t) \hat{\tilde{S}}_x {+} f_y^{\text{RF}}(t) \hat{\tilde{S}}_y \right]\! \left[ \hat{\tilde{I}}^{(i)}_x \cos(\omega_c t){+} \hat{\tilde{I}}^{(i)}_y \sin(\omega_c t) \right]\bigg\}. \label{eq:H_avg}
\end{multline}
Here, various terms are excluded as they time average to zero (rotating-wave approximation). The $\sqrt{7}$ prefactor comes from mapping the original spin-7/2 operators to the effective spin-1/2 ones. Additionally, the energy shift induced by $B^{\text{OH}}_z$ and time-dependent $B^{\text{RF}}$ is cancelled since we are using square-wave RF. 
The Fourier transforms of the modulation functions $f_{x,y}(t)$, termed the filter functions \cite{Degen2017}, directly reveal resonance frequencies at which equation~(\ref{eq:H_avg}) yields non-zero contributions (Extended~Data~Fig.~4b). Resonant interactions with strength proportional to the nuclear Overhauser field are achieved at sequence periods $2\tau$ which satisfy $\frac{1}{2\tau} = \frac{\omega_c}{2\pi \times 2}, \frac{\omega_c}{2\pi \times 4}, \frac{\omega_c}{2\pi \times 6}, \cdots$; interactions proportional to the RF field occur at sequence periods satisfying $\frac{1}{2\tau} = \frac{\omega_c}{2\pi \times 1}, \frac{\omega_c}{2\pi \times 3}, \frac{\omega_c}{2\pi \times 5}, \cdots$. Critically, these two sets of resonances occur at different values of $2\tau$, hence we can preferentially utilise the coherent, RF-induced interactions whilst decoupling from those induced by the randomised Overhauser field. This is experimentally demonstrated in Fig.~2b where the RF-induced resonances are spectrally resolved. In this measurement the linewidth of the register resonances are limited by that of the filter function. We also note that the $\omega_a$ transition cannot be independently addressed by the ZenPol sequence due to the multiplicity of the three \textsuperscript{51}V transitions determined by the quadratic Hamiltonian ($\omega_a=\omega_b/2=\omega_c/3$).

We use the RF-driven resonance identified at $\frac{1}{2\tau} = \frac{\omega_c}{2\pi \times 5}$ by setting the free-evolution interval to $\frac{\tau}{4} = \frac{5\pi}{4\omega_c}$.
Under this resonance condition, the average Hamiltonian (equation~\eqref{eq:H_avg}) is simplified to  
\begin{align}
	\hat{H}_\text{avg} &=  -\sqrt{7}\left(\frac{1+\sqrt{2}}{5\pi}\right)a_{x}B^{\text{RF}} \times \nonumber \\
& \qquad \qquad	\sum_{i \in \text{register}} \left((\hat{\tilde{S}}_x + \hat{\tilde{S}}_y) \hat{\tilde{I}}^{(i)}_x + (- \hat{\tilde{S}}_x + \hat{\tilde{S}}_y )  \hat{\tilde{I}}^{(i)}_y \right) \nonumber \\
	&=  -\sqrt{7}\left(\frac{\sqrt{2}+2}{5\pi}\right) a_{x}B^{\text{RF}} \sum_{i \in \text{register}}\left(\hat{\tilde{S}}_x' \hat{\tilde{I}}^{(i)}_x + \hat{\tilde{S}}_y' \hat{\tilde{I}}^{(i)}_y\right) \nonumber \\
	&=b_{(5,\omega_c)}B^{\text{RF}} \sum_{i \in \text{register}}\left( \hat{\tilde{S}}_+' \hat{\tilde{I}}^{(i)}_- + \hat{\tilde{S}}_-' \hat{\tilde{I}}^{(i)}_+\right).
	\label{eq:H_ex}
\end{align}
Here, going from the first to the second line, we change the local $^{171}$Yb basis by rotating 45 degrees around the $z$-axis such that $\hat{\tilde{S}}_x' = (\hat{\tilde{S}}_x + \hat{\tilde{S}}_y)/\sqrt{2}, \hat{\tilde{S}}_y' = (-\hat{\tilde{S}}_x + \hat{\tilde{S}}_y)/\sqrt{2}$, and from the second to the third line, $\hat{\tilde{S}}_{\pm}' = \hat{\tilde{S}}_x' \pm i\hat{\tilde{S}}_y'$ and $\hat{\tilde{I}}_{\pm} = \hat{\tilde{I}}_x \pm i\hat{\tilde{I}}_y$ are used. We define the coefficient $b_{(k,\omega_j)}$ which determines the interaction strength for the $k^\text{th}$ resonance addressing transition $\omega_j$ (for example, $b_{(5,\omega_c)} = -\sqrt{7}(\sqrt{2}+2)a_x/10\pi$). In the main text, we omit the primes on the $^{171}$Yb qubit operators for the sake of notational simplicity. The same analysis can be performed for other transitions, yielding a similar spin-exchange Hamiltonian, albeit with different interaction strength.

\subsection*{Direct Drive Gates for $^{51}$V Register}
Performing dynamical decoupling on the register requires selective driving of the froze-core $^{51}$V nuclear spins without perturbing the bath and is achieved through a two-fold mechanism. Firstly we initialise the $^{171}$Yb qubit into $\ket{0_g}$ and apply a sinusoidal $z$-directed RF magnetic field at $\omega_c/2\pi = 991$~kHz through the coplanar waveguide to induce an oscillating $^{171}$Yb magnetic dipole moment (Extended~Data~Fig.~7a). This generates an $x$-directed field component at each $^{51}$V spin, where the driving Hamiltonian is given by $\hat{H}_\text{drive}=\mu_Ng_{vx}A_x B^\text{osc}_{z}\sin(\omega_c t)\Hat{I}_x$
with $A_x=-3ln \mu_0\gamma_z^2/8\pi r^3\omega_{01}$. Here, $\mu_N$ is the nuclear magneton, $g_{vx}$ is the  $^{51}$V $x$-directed $g$-factor, $B_z^\text{osc}$ is the sinusoidal RF magnetic field amplitude, $\hat{I}_x$ is the nuclear spin-7/2 operator along the $x$-axis, $\{l,n\}$ are the $\{x,z\}$ directional cosines of the $^{171}$Yb--$^{51}$V displacement vector, $\mu_0$ is the vacuum permittivity, and $r$ is the $^{171}$Yb--$^{51}$V ion distance (Supplementary Information). The lattice symmetry of the host leads to equidistant spacing of the four proximal $^{51}$V spins from the central $^{171}$Yb qubit allowing homogeneous coherent driving of all register spins. 

In this direct driving scheme, we note that the effect of $B_z^\text{osc}$ is amplified by a factor of $A_x \approx 6.7$ for the frozen-core register spins at a distance of r = 3.9~{\AA} (Supplementary Information). Crucially, the amplification factor scales as $A_x \propto 1/r^3$ with distance $r$ from the $^{171}$Yb qubit, leading to a reduced driving strength for distant $^{51}$V bath spins. Moreover, the transition frequency of the bath, $\omega_c^\text{bath}/2\pi = 1028$~kHz, is detuned by 37~kHz from that of the register, $\omega_c/2\pi = 991$~kHz, further weakening the bath interaction due to off-resonant driving provided that the Rabi frequency is less than the detuning. 

In a rotating frame at frequency $\omega_c$, the driving Hamiltonian $\hat{H}_\text{drive}$ gives rise to Rabi oscillation dynamics of the register spins within the $\omega_c$ manifold, $\{\ket{\uparrow} = \ket{\pm5/2}, \ket{\downarrow} = \ket{\pm7/2}\}$. To calibrate \textsuperscript{51}V $\pi$ pulse times, we initialise the register into $\ket{0_v} = \ket{\downarrow\downarrow\downarrow\downarrow}$, drive the register for variable time, and read out the $\ket{0_v}$ population by preparing the $^{171}$Yb qubit in $\ket{1_g}$ and applying a swap gate to the $\omega_c$ transition. If the final $^{51}$V spin state is in $\ket{\downarrow}$ ($\ket{\uparrow}$)  the swap will be successful (unsuccessful) and the $^{171}$Yb qubit will end up in $\ket{0_g}$ ($\ket{1_g}$). Using this method, we induce resonant Rabi oscillations of the register at a Rabi frequency of $2\pi\times(7.65\pm0.05)$ kHz (blue markers,  Extended~Data~Fig.~7c) which exhibit exponential decay on a $280\pm30~\mu$s timescale, limited by dephasing caused by the fluctuating $^{171}$Yb Knight field. This can be decoupled using motional narrowing techniques whereby we periodically apply $\pi$ pulses to the $^{171}$Yb every 6 $\mu s$ during the drive period. In order to drive the $^{51}$V spins in a phase-continuous manner, we compensate for the inversion of the $^{171}$Yb magnetic dipole moment after each $\pi$ pulse by applying a $\pi$ phase shift to the sinusoidal driving field (Extended~Data~Fig.~7b). This leads to an extended $1/e$ Gaussian decay time of $1040\pm70~\mu$s (red markers, Extended~Data~Fig.~7c).

The arrow in Extended~Data~Fig.~7c indicates the 69~$\mu$s \textsuperscript{51}V $\pi$ pulse time used for dynamical decoupling. In contrast to the spin-preserving exchange interaction, this direct drive protocol provides independent, local control of the four $^{171}$V spins with no constraints on the number of excitations, thereby coupling the $^{51}$V register to states outside the two-level manifold spanned by $\ket{0_v}$ and $\ket{W_v}$. For example, at odd multiple $\pi$ times, we find
\begin{align*}
&\ket{0_v}\rightarrow\Ket{\uparrow\uparrow\uparrow\uparrow}\\
&\ket{W_v}\rightarrow\frac{\left(\Ket{\downarrow\uparrow\uparrow\uparrow}+\Ket{\uparrow\downarrow\uparrow\uparrow}+\Ket{\uparrow\uparrow\downarrow\uparrow}+\Ket{\uparrow\uparrow\uparrow\downarrow}\right)}{2},
\end{align*}
both of which contain more than a single excitation. For this reason, we use an even number of \textsuperscript{51}V $\pi$ pulses in our decoupling sequences to always return the $^{51}$V register to the memory manifold prior to state retrieval.

\subsection*{Population Basis Measurements}
\label{readout}
We develop a sequential tomography protocol \cite{Kalb2017} to read out the populations of the joint $^{171}$Yb--$^{51}$V density matrix $\rho$ in the effective four-state basis, $\{\ket{0_g0_v}, \ket{0_gW_v}, \ket{1_g0_v}, \ket{1_gW_v}\}$. This is achieved using two separate sequences: Readout sequence 1 and Readout sequence 2, applied alternately, which measure the $\{\ket{0_g0_v}$, $\ket{0_gW_v}\}$ and $\{\ket{1_g0_v}$, $\ket{1_gW_v}\}$ populations respectively. As shown in Extended~Data~Fig.~9a, these sequences are distinguished by the presence (absence) of a single $\pi$ pulse applied to the $^{171}$Yb qubit at the start of the sequence. This is followed by a single optical readout cycle on the A transition; results are post-selected on detection of a single optical photon during this period. Hence the presence (absence) of the first $\pi$ pulse results in $\ket{0_g}$ ($\ket{1_g}$) state readout after post selection. Furthermore, in all post-selected cases the $^{171}$Yb qubit is initialised to $\ket{1_g}$ by taking into account this conditional measurement outcome. Subsequently, an unconditional $\pi$ pulse is applied to the $^{171}$Yb, preparing it in $\ket{0_g}$ and a swap gate is applied, thereby transferring the $^{51}$V state to the $^{171}$Yb. Finally, we perform single-shot readout of the $^{171}$Yb state according to the protocol developed in \cite{Kindem2020}. Specifically, we apply two sets of 100 readout cycles to the A transition separated by a single $\pi$ pulse which inverts the $^{171}$Yb qubit population. The $^{51}$V state is ascribed to $\ket{W_v}$ ($\ket{0_v}$) if $\geq 1$ (0) photons are detected in the second readout period and 0 ($\geq 1$) photons are detected in the third. We summarise the possible photon detection events and state attributions in Extended~Data~Fig.~9b.

We demonstrate this protocol by characterizing the state preparation fidelities of the four basis states. The measured histograms are presented in Extended~Data~Fig.~9c alongside the respective gate sequences used for state preparation. The resulting uncorrected (corrected) preparation fidelities for these four basis states are:
\begin{align}
\mathcal{F}_{\ket{0_g0_v}} = 0.79\pm0.01 \; (0.82\pm0.02), \nonumber \\ 
\mathcal{F}_{\ket{0_gW_v}} = 0.50\pm0.02 \; (0.64\pm0.02), \nonumber \\
\mathcal{F}_{\ket{1_g0_v}} = 0.79\pm0.01 \; (0.82\pm0.02), \nonumber \\
\mathcal{F}_{\ket{1_gW_v}} = 0.50\pm0.02 \; (0.64\pm0.02). \nonumber
\end{align} 
We note that the reduced fidelity of $\ket{0_gW_v}$ and $\ket{1_gW_v}$ relative to $\ket{0_g0_v}$ and $\ket{1_g0_v}$ arises from the swap gate used for the $\ket{W_v}$ state preparation. Finally, we also characterize the fidelity of the maximally entangled $^{171}$Yb--$^{51}$V Bell state, $\ket{\Psi^+}=\frac{1}{\sqrt{2}}\left(\ket{1_g0_v}-i\ket{0_gW_v}\right)$, prepared using a single $\sqrt{\text{swap}}$ gate as described in the main text (Extended~Data~Fig.~9d). The corresponding uncorrected (corrected) populations for the four basis states, denoted $p_{ij}$ ($c_{ij}$) are:
\begin{align}
p_{00} \equiv \bra{0_g0_v}\rho \ket{0_g0_v} &= 0.16\pm0.01 \; (c_{00}=0.07\pm0.02 ), \nonumber \\
p_{01} \equiv \bra{0_gW_v}\rho \ket{0_gW_v} &= 0.32\pm0.01 \; (c_{01}=0.41\pm0.02), \nonumber \\
p_{10} \equiv \bra{1_g0_v} \rho \ket{1_g0_v} &= 0.40\pm0.02 \; (c_{10}=0.41\pm0.02), \nonumber \\
p_{11} \equiv \bra{1_gW_v} \rho \ket{1_gW_v} &= 0.12\pm0.01 \; (c_{11}=0.11\pm0.01). \nonumber
\end{align}

\subsection*{Swap Gate Fidelity Correction}
\label{fidcor}
Since $^{171}$Yb readout fidelity is $>95\%$ \cite{Kindem2020}, the dominant error introduced during the population basis measurements arises from the swap gate. We measure its fidelity in the population basis by preparing either the $\ket{0_g0_v}$ state (zero spin excitations) or the $\ket{1_g0_v}$ state (single spin excitation) and applying two consecutive swap gates such that the system is returned to the initial state. By comparing the $^{171}$Yb population before ($p_\text{pre}$) and after ($p_\text{post}$) the two gates are applied, we can extract fidelity estimates independently from the $^{51}$V state initialisation. Assuming the swap and swap-back processes are symmetric, we obtain a gate fidelity $\mathcal{F}_\text{sw}=\sqrt{(1-2p_\text{post})/(1-2p_\text{pre})}$. This quantity is measured for zero spin excitations leading to $\mathcal{F}_{\text{sw}, 0}=0.83$ and with a single spin excitation leading to $\mathcal{F}_{\text{sw},1}=0.52$.

When measuring the joint $^{171}$Yb--$^{51}$V populations $\{p_{00}$, $p_{01}$, $p_{10}$, $p_{11}\}$ we can use these fidelities to extract a set of  corrected populations $\{c_{00}, c_{01}, c_{10}, c_{11}\}$ according to the method described in \cite{Bernien2013,Nguyen2019a} using
\begin{equation} \label{pcrel}
    \begin{pmatrix}c_{11}\\c_{10}\\c_{01}\\c_{00}\end{pmatrix}=E^{-1}\begin{pmatrix}p_{11}\\p_{10}\\p_{01}\\p_{00}\end{pmatrix},
\end{equation}
where
\begin{align*}
    E {=} \frac{1}{2}\begin{pmatrix}1{+}\mathcal{F}_\text{sw,1}&&1{-}\mathcal{F}_\text{sw,0}&&0&&0\\1{-}\mathcal{F}_\text{sw,1}&&1{+}\mathcal{F}_\text{sw,0}&&0&&0\\0&&0&&1+\mathcal{F}_\text{sw,1}&&1{-}\mathcal{F}_\text{sw,0}\\0&&0&&1-\mathcal{F}_\text{sw,1}&&1{+}\mathcal{F}_\text{sw,0}\end{pmatrix}.
\end{align*}
We use a similar approach to correct the $\sqrt{\text{swap}}$ gate used to read out the Bell state coherence (Supplementary Information).
\clearpage
\onecolumngrid
\renewcommand{\theequation}{S\arabic{equation}}
\setcounter{equation}{0}
\section*{Supplementary Information}
\subsection{$^{171}$Y\MakeLowercase{b}--$^{51}$V Interactions}\label{hamderiv}
\subsubsection{Ground State $^{171}$Yb Hamiltonian}
The effective spin-1/2 Hamiltonian for the $^2\text{F}_{7/2}(0)$ $^{171}\text{Yb}^{3+}$ ground state is given by \cite{Kindem2018a}:
\begin{equation}
    \hat{H}_\text{eff}=\mu_B\mathbf{B}\cdot\mathbf{g}\cdot\Hat{\mathbf{S}}+\Hat{\mathbf{I}}_\text{Yb}\cdot\mathbf{A}\cdot\Hat{\mathbf{S}}
\end{equation}
where $\mathbf{B}$ is the magnetic field, $\Hat{\mathbf{S}}$ and $\Hat{\mathbf{I}}_\text{Yb}$ are vectors of $^{171}$Yb electron and nuclear spin-1/2 operators respectively and we neglect the nuclear Zeeman term.
The ground state $\mathbf{g}$ tensor is given by:
\begin{equation}
\mathbf{g}=
\begin{pmatrix}
    g_x & 0 & 0\\
    0 & g_x & 0\\
    0 & 0 & g_z
    \end{pmatrix}=
    \begin{pmatrix}
    0.85 & 0 & 0\\
    0 & 0.85 & 0\\
    0 & 0 & -6.08
    \end{pmatrix},
\end{equation}
which is a uniaxial tensor with the extraordinary axis parallel to the $c$-axis of the crystal and the two ordinary axes aligned with the crystal $a$-axes.
The ground state $\mathbf{A}$ tensor is given by:
\begin{equation}
\mathbf{A}= 2\pi \times
    \begin{pmatrix}
    0.675 & 0 & 0\\
    0 & 0.675 & 0\\
    0 & 0 & -4.82
    \end{pmatrix}
    \text{GHz}.
\end{equation}
Extended Data Fig.~1a shows the zero magnetic field energy level structure with hybridised $^{171}$Yb electron-nuclear spin eigenstates. Note that the zero-field $^{171}$Yb qubit states,  $\ket{0_g}$ and $\ket{1_g}$, have no magnetic dipole moment. See \cite{Kindem2018a} for more details. Throughout this work we adopt an $\hbar=1$ convention.

\subsubsection{Local Nuclear Spin Environment}
\label{localenv}
The \textsuperscript{171}Yb\textsuperscript{3+} ion substitutes for yttrium in a single site of the YVO$_4$ crystal, furthermore naturally abundant V and Y  contain 99.8\% \textsuperscript{51}V and 100\% \textsuperscript{89}Y isotopes, respectively. Hence each $^{171}$Yb ion experiences a near-identical nuclear spin environment.
The $^{51}$V ions have nuclear spin-7/2 leading to electric quadrupole interactions that cause a zero-field splitting. The resulting zero-field energy level structure of the \textsuperscript{51}V spins is given by:
\begin{equation}
    \hat{H}_{\text{V}}=Q \hat{I}_z^2
\end{equation}
where $Q/2\pi = 171~\text{kHz}$ measured using nuclear magnetic resonance (NMR) on bulk YVO\textsubscript{4} crystals \cite{Bleaney1982} and $\hat{I}_z$ is the \textsuperscript{51}V nuclear spin-7/2 spin operator along the $c\equiv z$ axis. Note that the local  \textsuperscript{51}V register ions surrounding the \textsuperscript{171}Yb qubit experience a frozen-core detuning as discussed in the main text, leading to a smaller quadrupolar splitting with $Q/2\pi = 165~\text{kHz}$. The energy level structure of these register ions is shown in Fig.~1b. The \textsuperscript{89}Y ion, on the other hand, has no zero-field structure.

The positions of the six nearest \textsuperscript{51}V ions are tabulated below, where $\mathbf{r}=[x~y~z]$ is the \textsuperscript{171}Yb--\textsuperscript{51}V position vector with magnitude $r$ and direction cosines $\{l,m,n\}$.

\begin{center}
\begin{tabular}{ |c|c|c|c|c|c|c|c|c| }
\hline
\textsuperscript{51}V ion \# & Shell & $r$ ($\text{\AA}$) & $x$ ($\text{\AA}$) & $y$ ($\text{\AA}$)& $z$ ($\text{\AA}$) & $l$ & $m$ & $n$\\
\hline
 1 & 1\textsuperscript{st} & 3.1 & 0 & 0 & -3.1 & 0 & 0 & -1 \\ 
 \hline
 2 & 1\textsuperscript{st} & 3.1 & 0 & 0 & 3.1 & 0 & 0 & 1 \\  
 \hline
 3 & 2\textsuperscript{nd} & 3.9 & 0 & -3.6 & 1.6 & 0 & -0.91 & 0.40 \\
 \hline
  4 & 2\textsuperscript{nd} & 3.9 & 0 & 3.6 & 1.6 & 0 & 0.91 & 0.40 \\
 \hline
 5 & 2\textsuperscript{nd} & 3.9 & -3.6 & 0 & -1.6 & -0.91 & 0 & -0.40 \\
 \hline
6 & 2\textsuperscript{nd} & 3.9 & 3.6 & 0 & -1.6 & 0.91 & 0 & -0.40 \\
 \hline
\end{tabular}
\end{center}
Note that the two nearest \textsuperscript{51}V ions (1 and 2) are located directly above and below the $^{171}$Yb qubit along the $z$-axis, due to their positions they cannot be driven by the induced $^{171}$Yb magnetic dipole moment and thus belong to the bath (Supplementary Information Section~\ref{bathreg}). In contrast, ions 3--6 are symmetrically positioned in the lattice with non-zero $x/y$ and $z$ coordinates, forming the frozen-core register spins utilized as a quantum memory. 

The $^{51}$V ions have a uniaxial g-tensor with form \cite{Bleaney1982a}:
\begin{equation}
\mathbf{g}_\text{V}=
\begin{pmatrix}
    g_{vx} & 0 & 0\\
    0 & g_{vx} & 0\\
    0 & 0 & g_{vz}
    \end{pmatrix}
\end{equation}
See Supplementary Information Section~\ref{simulations} for an experimental estimation of these tensor components.

\subsubsection{$^{171}$Yb--$^{51}$V Interactions}
\label{interactions}
The magnetic dipole-dipole interaction between the $^{171}$Yb qubit and a single $^{51}$V ion can be described by the following Hamiltonian:
\begin{equation}
    \hat{H}_{dd}=\frac{\mu_0}{4\pi}\left[\frac{\boldsymbol{\mu}_\text{Yb}\cdot\boldsymbol{\mu}_\text{V}}{r^3}-\frac{3(\boldsymbol{\mu}_\text{Yb}\cdot\mathbf{r})(\boldsymbol{\mu}_\text{V}\cdot\mathbf{r})}{r^5}\right]
\end{equation}
where $\boldsymbol{\mu}_\text{Yb}=-\mu_B\mathbf{g}\cdot\Hat{\mathbf{S}}$, $\boldsymbol{\mu}_\text{V}=\mu_N\mathbf{g}_\text{V}\cdot\Hat{\mathbf{I}}$ (note that $\hat{\bf S}$ and $\hat{\bf I}$ are vectors of $^{171}$Yb and $^{51}$V spin operators, respectively), $\mu_B$ is the Bohr magneton, $\mu_N$ is the nuclear magneton, $\mu_0$ is the vacuum permeability and $\mathbf{r}$ is the $^{171}$Yb--$^{51}$V displacement vector with magnitude $r$. Due to the highly off-resonant nature of the $^{171}$Yb--$^{51}$V interaction, a secular approximation would be appropriate. To first order, however, all secular terms involving the $^{171}$Yb qubit basis are zero, i.e., $\bra{0_g}\hat{H}_{dd}\ket{0_g}=0$, $\bra{1_g}\hat{H}_{dd}\ket{1_g}=0$.

To proceed, we consider second-order effects which generally scale as $\sim$$g^2/\Delta E$, where $\Delta E$ is the energy separation between a pair of unperturbed eigenstates. By taking into account the fact that $g_z$ is roughly 7 times larger than $g_x, g_y$ and $\Hat{S_z}$ terms in $\hat{H}_{dd}$ mix $\ket{0_g}$ and $\ket{1_g}$ with small $\Delta E$ whereas $\Hat{S_x}$ and $\Hat{S_y}$ mix the $^{171}$Yb qubit states and $\ket{\text{aux}_g}$ with large $\Delta E$, we restrict our consideration to the $\Hat{S}_z$ terms in $\hat{H}_{dd}$:
\begin{equation}
    \hat{H}_{dd}\approx\frac{\mu_0\mu_B\mu_Ng_z}{4\pi r^3}\hat{S}_z\left[3lng_{vx}\hat{I}_x+3mng_{vx}\hat{I}_y+(3n^2-1)g_{vz}\hat{I}_z\right] \label{eq:H_dd}
\end{equation}
where $\{l,m,n\}$ are direction cosines of the $^{171}$Yb--$^{51}$V displacement vector. Note that the $\hat{S}_z$ operator is the {\it electron} spin-1/2 operator defined as $\hat{S}_z = 1/2 (\ket{0_g}\bra{1_g} + \ket{1_g}\bra{0_g})$ in the basis of the hybridised eigenstates of the $^{171}$Yb qubit.

\subsubsection{Nuclear Overhauser Field}
\label{bathreg}
As discussed in the main text, we can divide the \textsuperscript{51}V spins into two ensembles: register spins and bath spins. The bath spins comprise $^{51}$V ions which are not driven by the $^{171}$Yb qubit for the following two reasons:
\begin{enumerate}
    \item \textbf{Ions which aren't driven due to position}: certain ions (such as 1 and 2 in the above table) only interact via an Ising-type $\Hat{S}_z\Hat{I}_z$ Hamiltonian. Hence the $^{171}$Yb qubit cannot be used to drive transitions between the $^{51}$V $z$-quantised quadrupole levels.
    \item \textbf{Ions which aren't driven due to detuning}: As observed in the ZenPol spectra (Fig.~2b in the main text), more distant spins are spectrally separated from the nearby ions comprising the register.
\end{enumerate}
We assume that the bath spins are in an infinite-temperature mixed state: $\rho_\text{V}=\mathbb{1}_\text{V}/Tr\{\mathbb{1}_\text{V}\}$, where $\mathbb{1}_v$ is the identity matrix in the Hilbert space for the bath spins.  In the mean field picture, their effect on the $^{171}$Yb can be approximated as a classical fluctuating magnetic field, commonly termed the nuclear Overhauser field. As mentioned previously, since $g_z^2 \gg g_{x,y}^2$, the $z$-component of the Overhauser field is dominant, given by
\begin{equation}
\label{overhauser}
    B^\text{OH}_z=\sum_{\text{i}\in\text{bath}}\frac{\mu_0\mu_Ng_{vz}}{4\pi (r^{(i)})^3}(3(n^{(i)})^2-1)m_I^{(i)},
\end{equation}
where $r^{(i)}$ and $n^{(i)}$ are the distance and $z$-direction cosine between the \textsuperscript{171}Yb and $i$\textsuperscript{th} bath spin, and $m_I^{(i)} \in \{-7/2, -5/2, ..., 5/2, 7/2\}$ is the nuclear spin projection at site $i$. Note that $B_z^\text{OH}$ is randomly fluctuating due to the stochastic occupation of the 8 possible $\ket{m_I}$ states, however, it is quasi-static on the timescale of our control sequences, hence we do not label the time dependence.

Crucially, the nuclear Overhauser field generates some weak mixing between $\ket{0_g}$ and $\ket{1_g}$ leading to perturbed eigenstates $\ket{\tilde{0}_g}$ and $\ket{\tilde{1}_g}$ which have a small, induced, $z$-directed dipole moment. These states have the form
\begin{align}
    \ket{\tilde{0}_g}&=\ket{0_g}-\frac{\gamma_z (B^\text{OH}_z+B^\text{RF}(t))}{2\omega_{01}}\ket{1_g} \nonumber\\
    \ket{\tilde{1}_g}&=\ket{1_g}+\frac{\gamma_z (B^\text{OH}_z+B^\text{RF}(t))}{2\omega_{01}}\ket{0_g}
\end{align}
where $\gamma_z=g_z\mu_B$ is the longitudinal gyromagnetic ratio of the $^{171}$Yb qubit and $\omega_{01}/2\pi = 675$~MHz is the unperturbed $^{171}$Yb $\ket{0_g}\leftrightarrow\ket{1_g}$ transition frequency. Here we have added the effect of an externally applied, $z$-directed, square-wave RF magnetic field $B^\text{RF}(t)$ with amplitude $B^\text{RF}$ used in the ZenPol sequence (see main text for details); note that this field is piecewise constant for each half-sequence period, hence the time dependence corresponds to periodic flips between $\pm B^\text{RF}$. In addition, these fields induce a detuning of the $^{171}$Yb $\ket{0_g}\leftrightarrow\ket{1_g}$ transition, which can be calculated using second-order perturbation theory as $\Delta(t)= \gamma_z^2 (B^\text{OH}_{z}+B^\text{RF}(t))^2/2\omega_{01}$.

\subsubsection{Interaction with Register Ions}
\label{regint}
We postulate that the second nearest shell of four $^{51}$V ions (ions 3--6 in the table above) comprise the register. These four ions are equidistant from the $^{171}$Yb and interact via both an $\Hat{S}_z\Hat{I}_z$ term and $\Hat{S}_z\Hat{I}_x$ or $\Hat{S}_z\Hat{I}_y$ terms. To identify an effective  interaction Hamiltonian in the perturbed basis $\{\ket{\tilde{0}_g}, \ket{\tilde{1}_g}\}$, we consider only secular matrix elements of $\hat{H}_{dd}$ (equation~\eqref{eq:H_dd}):
\begin{equation}
    \hat{\tilde{H}}_{dd} =\bra{\tilde{0}_g}\hat{H}_{dd}\ket{\tilde{0}_g}\ket{\tilde{0}_g}\bra{\tilde{0}_g}+\bra{\tilde{1}_g}\hat{H}_{dd}\ket{\tilde{1}}\ket{\tilde{1}_g}\bra{\tilde{1}_g}
\end{equation}
where
\begin{align}
    \bra{\tilde{0}_g}\hat{H}_{dd}\ket{\tilde{0}_g} &=-\frac{\mu_0 \mu_N \gamma_z^2 (B^{\text{OH}}_z+B^\text{RF}(t))}{8\pi r^3\omega_{01}}\left[3lng_{vx}\hat{I}_x+3mng_{vx}\hat{I}_y+(3n^2-1)g_{vz}\hat{I}_z\right] \nonumber \\
    \bra{\tilde{1}_g}\hat{H}_{dd}\ket{\tilde{1}_g} &=+\frac{\mu_0\mu_N\gamma_z^2(B^{\text{OH}}_z+B^\text{RF}(t))}{8\pi r^3\omega_{01}}\left[3lng_{vx}\hat{I}_x+3mng_{vx}\hat{I}_y+(3n^2-1)g_{vz}\hat{I}_z\right]. \nonumber
\end{align}
Hence the effective interaction between the $^{171}$Yb qubit and the four register spins, $\hat{H}_\text{int} = \sum_{i \in \text{register}} \hat{\tilde{H}}_{dd}^\text{(i)}$, can be described by
\begin{equation}
\hat{H}_\text{int}=\hat{\Tilde{S}}_z(B^\text{OH}_z+B^\text{RF}(t))\sum_{i\in\text{register}}\left(J_x^{(i)} \hat{I}_x^{(i)}+J_y^{(i)} \hat{I}_y^{(i)}+J_z^{(i)} \hat{I}_z^{(i)}\right)
\end{equation}
with
\begin{align*}
J_x^{(i)}&=\frac{3\mu_0\mu_N\gamma_z^2 g_{vx}l^{(i)}n^{(i)}}{4\pi (r^{(i)})^3\omega_{01}}\\
J_y^{(i)}&=\frac{3\mu_0\mu_N\gamma_z^2 g_{vx}m^{(i)}n^{(i)}}{4\pi (r^{(i)})^3\omega_{01}}\\
J_z^{(i)}&=\frac{3\mu_0\mu_N\gamma_z^2 g_{vz}(3(n^{(i)})^2-1)}{4\pi (r^{(i)})^3\omega_{01}}
\end{align*}
and
\begin{equation*}
    \hat{\tilde{S}}_z=\frac{1}{2}(\ket{\tilde{1}_g}\bra{\tilde{1}_g}-\ket{\tilde{0}_g}\bra{\tilde{0}_g}).
\end{equation*}
Finally, we perform local basis transformations of each $^{51}$V ion to further simplify the Hamiltonian form. Specifically, we apply the following unitary rotation:
\begin{align*}
    \hat{H}_\text{int}& \rightarrow U \hat{H}_\text{int}U^\dag \nonumber \\
    U&=\prod_{j\in\text{register}}\exp[i\theta^{(j)}\hat{I}_z^{(j)}],
\end{align*}
where $\theta^{(j)}= \tan^{-1}(m^{(j)}/l^{(j)})$, which leads to
\begin{equation}
\hat{H}_\text{int}=\hat{\tilde{S}}_z(B^\text{OH}_z+B^\text{RF}(t))\sum_{i\in\text{register}}\left[a_{x}\hat{I}_x^{(i)}+a_z \hat{I}_z^{(i)}\right]
\end{equation}
with $a_x = \sqrt{(J_x^{(i)})^2+(J_y^{(i)})^2}$ and $a_z=J_z^{(i)}$. Note that the coupling coefficients $a_x$ and $a_z$ are homogeneous (i.e. independent of site index $i$) since the four register spins are equidistant from the central $^{171}$Yb and have directional cosine factors with equal magnitude. 

The same result can also be derived using the Schrieffer-Wolff transformation \cite{Tannoudji2004,Bermudez2011}, where the interaction Hamiltonian obtained here corresponds to the dominant second-order perturbation terms. Hereafter we simplify our notation and use $\ket{0_g}$ and $\ket{1_g}$ without tildes to represent the weakly perturbed eigenstates in the presence of any small magnetic field.

\subsubsection{Full System Hamiltonian}
\label{fullham}
Combining the various energy and interaction terms, the full system Hamiltonian (in a $^{171}$Yb frame rotating at $\omega_{01}/2\pi = 675$~MHz) becomes:
\begin{equation}
    \hat{H}_\text{full} =\frac{\gamma_z^2\left(B^\text{OH}_z+B^\text{RF}(t)\right)^2}{2\omega_{01}}\hat{\tilde{S}}_{z}+\sum_{i\in\text{register}} Q\left(\Hat{I}_z^{(i)}\right)^2+\hat{\tilde{S}}_z(B^\text{OH}_z+B^\text{RF}(t))\sum_{i\in\text{register}}\left[a_{x}\hat{I}_x^{(i)}+a_{z}\hat{I}_z^{(i)}\right].
\end{equation}

\subsection{Randomised Benchmarking and \textsuperscript{171}Y\MakeLowercase{b} Qubit Coherence}
High fidelity control of the $^{171}$Yb $\ket{0_g}\leftrightarrow\ket{1_g}$ transition is essential for implementing the ZenPol sequence and enabling coherent $^{171}$Yb--$^{51}$V interactions. For example, a single swap operation realised by the ZenPol sequence contains 120 local $^{171}$Yb gates. We characterise our single qubit gate fidelity using randomised benchmarking \cite{Knill2008}, which provides a value independent from state preparation or measurement (SPAM) errors. We apply randomly sampled single qubit Clifford gates constructed using $\pi$ and $\pi/2$ rotations around the $x$ and $y$ directions followed by the single-gate inverse operation (Extended~Data~Fig.~2a). When the number of gates, $M_\text{gate}$, increases, the sequence error accumulates and the probability of returning to the initial $\ket{0_g}$ state reduces according to an exponential decay:
\begin{equation}
P=0.5+P_0d^{M_\text{gate}}.
\end{equation}
When ensemble-averaged over a sufficiently large number of random gate sets (in our case 100), $f=\frac{1}{2}(1+d)$ becomes a reliable estimate of the average single-qubit gate fidelity. Measurement results are presented in Extended~Data~Fig.~2a, leading to an extracted average single qubit gate fidelity of $f = 0.99975\pm0.00004$.

We also measure the $T_2$ coherence time of the qubit transition using an XY-8 dynamical decoupling sequence \cite{Gullion1990}. Specifically, we work with a fixed inter-pulse separation of 5.6$~\mu$s and measure the coherence time by varying the number of decoupling periods, $M'$ (Extended~Data~Fig.~2b). We measure an exponential decay with $1/e$ time constant $T_2=16\pm2$~ms. We note that this measurement uses the same method as in \cite{Kindem2020}, however, we observe a factor of three improvement in coherence due to the improved microwave setup leading to correspondingly increased $\pi$ gate fidelities.

\subsection{Extra Register Detail}
In this section we provide additional technical details related to the single excitation states used to store quantum information on the $^{51}$V spins.

The general form for the engineered spin-exchange interaction is:
\begin{equation}
	\hat{H}_\text{avg} = B^\text{RF} \sum_{i \in \text{register}} b_{(k,\omega_j)}^{(i)}\left(\hat{\tilde{S}}_+ \hat{\tilde{I}}^{(i)}_- + \hat{\tilde{S}}_-\hat{\tilde{I}}^{(i)}_+\right),
\end{equation}
where, $B^\text{RF}$ is the square-wave RF magnetic field amplitude, $b_{(k,\omega_j)}^{(i)}$ is the $k^\text{th}$ resonance prefactor for transition $\omega_j$ of the $i$\textsuperscript{th} register spin, $\Hat{\tilde{I}}_+ = \ket{\uparrow} \bra{\downarrow}$, $\Hat{\tilde{I}}_- = \ket{\downarrow} \bra{\uparrow}$, are  raising and lowering operators in an effective nuclear spin-1/2 manifold and $\Hat{\tilde{S}}_+ = \ket{1_g} \bra{0_g}$, $\Hat{\tilde{S}}_- = \ket{0_g} \bra{1_g}$ are raising and lowering operators for the $^{171}$Yb qubit. Note, unlike the main text, we do not assume homogeneous coupling to the register spins, hence the $b_{(k,\omega_j)}^{(i)}$ coefficients depend on the register site index $i$. In addition, we consider an arbitrary number of register spins, $N$, that are spectrally indistinguishable.

When the \textsuperscript{171}Yb is initialised in $\ket{1_g}$ and the \textsuperscript{51}V register spins are polarised in $\ket{0_v}=\ket{\downarrow}^{\otimes N}$, this interaction leads to the following spin-exchange evolution \cite{Taylor2003}:
\begin{equation}
\ket{\psi(t)} = \ket{1_g} \ket{0_v} \cos(J_\text{ex} t/2)-i \ket{0_g} \ket{1_v} \sin(J_\text{ex} t/2)
\end{equation}
where the spin-exchange frequency is given by:
\begin{equation}
J_\text{ex}=2B^\text{RF}\sqrt{\sum_{i\in\text{register}}\left| b_{(k,\omega_j)}^{(i)}\right|^2}
\end{equation}
and the resulting single-spin excited state generated by this interaction is:
\begin{equation}
    \ket{1_v}=\frac{1}{\sqrt{\sum_i|b_{(k,\omega_j)}^{(i)}|^2}}\sum_{i \in \text{register}}b_{(k,\omega_j)}^{(i)}\ket{\downarrow...\uparrow^{(i)}...\downarrow}.
\end{equation}

Based on the results presented in Fig.~2d and Supplementary Information Section~\ref{singlespin} we postulate that for our system the register consists of the second nearest shell of four homogeneously coupled $^{51}$V ions.
In this case we recover the expressions presented in the main text, namely, the single-spin excitation in the register realises an entangled four-body W-state, $\ket{W_v}$, as depicted in Fig.~1c:
\begin{equation}
    \ket{1_v}=\ket{W_v}=\frac{\ket{\uparrow\downarrow\downarrow\downarrow}+\ket{\downarrow\uparrow\downarrow\downarrow}+\ket{\downarrow\downarrow\uparrow\downarrow}+\ket{\downarrow\downarrow\downarrow\uparrow}}{2}
\end{equation}
and the spin-exchange rate is given by $J_\text{ex}=4B^\text{RF}b_{(k,\omega_j)}$. In general, for $N$ homogeneously coupled register spins, we expect that the spin-exchange rate is enhanced by a factor of $\sqrt{N}$, leading to faster swap gate operation. 

We note that in this protocol it is possible to transfer a second spin excitation to the register. More specifically, the spin-preserving exchange interaction, $\hat{\tilde{S}}_-\hat{\tilde{I}}_+ + \hat{\tilde{S}}_+\hat{\tilde{I}}_-$, couples the state $\ket{1_g}\ket{W_v}$ to $\ket{0_g}\ket{2_v}$, where $\ket{2_v}$ is a $^{51}$V state with two spins in $\ket{\uparrow}$. To avoid undesired excitation to states outside of the effective $\{\ket{0_v},\ket{W_v}\}$ manifold, we always prepare the $^{171}$Yb qubit in $\ket{0_g}$ before retrieving stored states from the $^{51}$V register. Hence the swap gate realised by this interaction operates on a limited basis of states.

We stress that utilising the dense, lattice nuclear spins ensures near identical registers for all $^{171}$Yb ions. 
Extended Data Figure 10 shows ZenPol spectra near the $\omega_c$ transition, collectively enhanced spin-exchange oscillations and motionally-narrowed $T_2^*$ times for three \textsuperscript{51}V registers coupled to three different \textsuperscript{171}Yb ions. The $^{171}$Yb optical and microwave frequencies were re-calibrated for each ion, however, all aspects of the experimental sequences related to register control and readout were identical.

\subsection{Simulation} \label{simulations}
We simulate our coupled spin system using the effective Hamiltonian derived in Supplementary Information Section~\ref{fullham}, however we add three additional terms:
\begin{enumerate}
\item
{\bf Nuclear Zeeman interactions of the $^{51}$V register spins with the Overhauser field from the bath}: Since the energy levels are quantised along the $z$-axis, magnetic fluctuations along the $z$-direction dominate, which can be captured by the following Hamiltonian
\begin{equation}
\hat{H}_\text{nz}= \sum_{i \in \text{register}} \mu_Ng_{vz}B^{\text{OH}}_{z}\left(\mathbf{r}_i\right)\hat{I}_z^{(i)} \label{eq:Zeeman}
\end{equation}
where $B^{\text{OH}}_{z}\left(\mathbf{r}_i\right)$ is the $z$-component of the Overhauser field evaluated at the position of the $i$\textsuperscript{th} register ion, $\mathbf{r}_i$.
\item
{\bf Nuclear magnetic dipole-dipole interactions of the register spins}: 
\begin{equation}
\hat{H}_\text{ndd}=\sum\limits_{\substack{i,j\in\text{register}\\ i<j}} \frac{\mu_0}{4\pi}\left[\frac{\boldsymbol{\mu}_{V}^{(i)}\cdot\boldsymbol{\mu}_{V}^{(j)}}{r_{ij}^3}-\frac{3(\boldsymbol{\mu}_{V}^{(i)}\cdot\mathbf{r}_{ij})(\boldsymbol{\mu}_V^{(j)}\cdot\mathbf{r}_{ij})}{r_{ij}^5}\right]
\end{equation}
with $\mathbf{r}_{ij}$ the displacement vector between $^{51}$V register spins at sites $i$ and $j$.
\item
{\bf $^{171}$Yb-enhanced register spin-spin interactions}: These terms are derived by considering second-order perturbations using the Schrieffer-Wolff transformation \cite{Tannoudji2004,Bermudez2011}. For example, the dominant Ising-type terms take the form
\begin{equation}
\hat{H}_\text{edd}=\sum_{i,j\in\text{register}}\frac{1}{2\omega_{01}}\left[\left(3n^2-1\right)\frac{\mu_0\mu_N \gamma_z g_{vz}}{4\pi r^3}\right]^2\hat{\tilde{S}}_z\hat{I}^{(i)}_z\hat{I}^{(j)}_z,
\end{equation}
where $r$ and $n$ are the magnitude and $z$-direction cosine of the $^{171}$Yb--$^{51}$V register ion displacement vector. However, we note that the ZenPol sequence cancels these interactions to first order.
\end{enumerate}

By simulating $^{171}$Yb Ramsey coherence times we extract $g_{vz}\approx1.6$. We note that estimation of the bare \textsuperscript{51}V coherence time indicates a potential discrepancy in this value by up to $25\%$, discussed further in Supplementary Information Section~\ref{coherence}, however, this has a negligible impact on the ZenPol sequence simulations. We obtain an estimate for $g_{vx}\approx0.6$ by calibrating the RF field amplitude and comparing with the experimental results of direct $^{51}$V spin driving in Extended~Data~Fig.~7.

We compute the nuclear Overhauser field $B_z^\text{OH}$ according to equation~\eqref{overhauser} by randomly sampling the bath states for each Monte-Carlo simulation repetition. We include a simple model of the bath dynamics by incorporating stochastic jumps of the bath spins on magnetic-dipole allowed transitions.

We simulate the register spin dynamics in a reduced Hilbert space by considering only the $\omega_c$ manifold. This enables fast simulation of all four register spins plus the $^{171}$Yb qubit transition (Hilbert space with dimension 32). Imperfect polarisation of the $^{51}$V register into $\ket{\downarrow} = \ket{\pm 7/2}$ is categorised into two distinct types:
\begin{enumerate}
\item
Imperfect polarisation within the $\omega_c$ transition i.e. a small residual population $\epsilon_1$ in $\ket{\uparrow} = \ket{\pm5/2}$.
\item
Imperfect polarisation outside the $\omega_c$ manifold i.e. a small residual population $\epsilon_2$ in $\ket{\pm1/2}$ and $\ket{\pm3/2}$.
\end{enumerate}
This leads to a $\ket{\downarrow}$ population of $1-\epsilon_1-\epsilon_2$.
We incorporate incomplete polarisation by sampling different register initial states for each Monte-Carlo repetition. For case 1, this involves occasionally initialising a given $^{51}$V ion into $\ket{\uparrow}$, while for case 2 this involves reducing the Hilbert space dimension by removing the $^{51}$V ion from the simulation. We also take into account finite pulse duration effects by modeling the ZenPol sequence using 25~ns $\pi/2$ and 50~ns $\pi$ pulses (Extended~Data~Fig.~6a).

As shown in Extended~Data~Fig.~6d, the spin-exchange oscillations from numerical simulation (red dashed line) exhibit slower decay than the measured experimental results (red markers). We add a phenomenological exponential decay envelope, $ce^{-M/\tau_M}$, to the simulation results where $c$ and $\tau_M$ are free parameters, and $M$ is the ZenPol sequence period. The additional decay could be caused by heating due to the RF field, excess $^{171}$Yb dephasing or additional register spin interactions which we haven’t considered here. We fit this model by optimising multiple parameters: $\epsilon_1$, $\epsilon_2$, $B^\text{RF}$, $c$ and $\tau_M$. The resulting values of $\epsilon_1$ and $\epsilon_2$ are 0.12 and 0.04, respectively, indicating $\approx84\%$ polarisation into $\ket{\downarrow}$; the RF magnetic field amplitude is $B^\text{RF} \approx1.6$~G and the phenomenological exponential decay parameters are $c=0.8$ and $\tau_M=90$ leading to a close fit with the experimental results (red solid line, Fig.~2c and Extended~Data~Fig.~6d). Additional simulation results following this methodology with varying $B^\text{RF}$ and $\tau$ are presented in Extended~Data~Fig.~6.

Finally, we model the results with a single-spin excitation in the $\omega_c$-manifold by including the $\ket{\pm3/2}$ level in the simulation (Fig.~2d and Supplementary Information Section~\ref{singlespin}). The initial state used in this simulation is partially polarised between the $\ket{\pm3/2}$ level with population $1-\epsilon$ and the $\ket{\pm1/2}$ level with population $\epsilon$.  We use the same value of $B^\text{RF}=1.6$~G as in Fig.~2c, and optimise the polarisation level leading to $1-\epsilon=0.8$. The close correspondence between the measured and simulated oscillation profiles suggests that the register consists of the second shell of four homogeneously coupled $^{51}$V ions.

\subsection{Hartmann Hahn Spectroscopy}
In addition to the ZenPol spectra discussed in the main text, we use Hartmann-Hahn (HH) double resonance \cite{Hartmann1962} to perform spectroscopy of the nuclear spin environment. This method enables spin exchange between two systems with different transition frequencies by resonantly driving a qubit with a Rabi frequency that matches the energy level splitting of the environmental nuclear spins. In our case, we resonantly drive the $^{171}$Yb at 675 MHz to generate a pair of dressed states $\ket{\pm}=\frac{1}{\sqrt{2}}\left(\ket{0_g}\pm i\ket{1_g}\right)$ with splitting $\Omega$ which we sweep over a range $\approx$ $2\pi\times$(0--2.3) MHz (Extended~Data~Fig.~3). The $^{171}$Yb qubit is initialised into the $\ket{-}$ dressed state by a $\pi/2$ pulse preceding the driving period. If resonant with a nuclear spin transition, the $^{171}$Yb qubit undergoes spin exchange at a rate dictated by the interaction strength. Finally we read out the $^{171}$Yb $\ket{+}$ dressed state population to determine whether spin exchange has occurred.

Extended~Data~Fig.~3b shows experimental results of HH spectroscopy where we vary both the HH drive Rabi frequency ($\Omega$) and also the HH pulse duration ($t$). The counts plotted on the colour-bar are proportional to the $\ket{+}$ dressed state population. We find three clear resonances at evenly spaced pulse amplitudes 0.15, 0.30 and 0.45 corresponding to the $\omega_a$, $\omega_b$ and $\omega_c$ $^{51}$V transitions; notably, unlike ZenPol, the HH sequence only has one harmonic leading to a single resonant interaction per transition. Also note the lack of oscillations when varying the pulse duration, $t$, on resonance with either of the three transitions: this is because the spin exchange is driven by the randomised, Overhauser field induced $^{171}$Yb dipole moment. For this reason, the HH sequence cannot be used to generate the coherent exchange interaction necessary to realise a swap gate for our system. In the case of no driving ($\Omega = 0$), the signal rapidly saturates as $t$ increases as a result of Ramsey dephasing of the initial state. However, as $\Omega$ exceeds the $^{171}$Yb spin linewidth ($\sim50$ kHz \cite{Kindem2020}), this effect diminishes due to the emergence of spin-locking effects and consequently leads to an increased saturation timescale when not resonant with the \textsuperscript{51}V transitions. The resolution of this measurement is also limited by the $^{171}$Yb spin linewidth, and we therefore cannot resolve the split-resonance structure observed in the ZenPol spectra. The results agree well with simulations (Extended~Data~Fig.~3c) indicating that interactions with the $^{51}$V quadrupolar structure dominate these measurements.

\subsection{Polarisation of multi-level register nuclear spins}
Polarisation dynamics are explored using the PROPI method (polarisation readout by polarisation inversion) \cite{Scheuer2017}. This sequence uses the back-action of the $^{51}$V spins on the $^{171}$Yb to measure the register polarisation after successive ZenPol polarisation cycles. For instance, when polarising into $\ket{\uparrow} = \ket{\pm5/2}$ on the $\omega_c$ transition, the $^{171}$Yb is initialised into $\ket{1_g}$ and undergoes spin exchange with any $^{51}$V population in $\ket{\downarrow} = \ket{\pm7/2}$. The $^{171}$Yb $\ket{0_g}$  population after interaction is therefore related to the residual $^{51}$V $\ket{\downarrow}$ population. As presented in Extended~Data~Fig.~5a, we measure the $^{171}$Yb population after each of 20 consecutive polarisation cycles and observe a saturation after 10 cycles, indicating that the $^{171}$Yb polarisation has been transferred to the $^{51}$V register. The high-contrast signal obtained in this measurement is enabled by alternating the $^{51}$V polarisation direction, i.e. periods of polarisation into $\ket{\uparrow}$ are interleaved with periods of polarisation into $\ket{\downarrow}$. This mitigates the need to wait for slow register thermalisation ($T_1^{(0)}=0.54~s$, see Supplementary Information Section~\ref{T1decay}) between consecutive experiment repetitions. These measurements are repeated with ZenPol sequences on the $\omega_b$ transition, demonstrating similar levels of polarisation saturation after approximately 10 cycles (Extended~Data~Fig.~5b).

We also demonstrate the effect of incomplete register polarisation on the spin-exchange oscillation by varying the number of polarisation cycles on the $\omega_b$ and $\omega_c$ transitions before each experiment (Extended~Data~Fig.~5c). As expected, we see that coherent spin-exchange oscillations emerge as an increasing number of polarisation cycles are applied.

These results inform the design of polarisation sequences used in subsequent single-spin excitation experiments where 40 polarisation cycles interleaved between the $\omega_b$ and $\omega_c$ transitions are sufficient to polarise the register into $\ket{0_v} = \ket{\downarrow\downarrow\downarrow\downarrow}$. Based on simulations discussed in Supplementary Information Section \ref{simulations} we estimate this protocol achieves $\approx84\%$ polarisation into the $\ket{0_v}$ state. Note that we don't use the ZenPol sequence to directly polarise the $\omega_a$ transition due to spectral overlap with $\omega_b$ and $\omega_c$ (Fig.~2b). We postulate that the high degree of polarisation can still be achieved even in the absence of direct $\omega_a$ transition control due to two factors:
\begin{enumerate}
\item
The thermalisation timescale of the $\omega_a$ transition is significantly shorter than the interrogation time. Specifically, our experiments typically run for several minutes whereas the $\omega_a$ thermalisation rate is likely similar to $T_1^{(0)}=0.54$~s. Thus, undesired population in the $\ket{\pm1/2}$ level can still pumped to $\ket{\pm7/2}$ once it relaxes to $\ket{\pm3/2}$.
\item
Once successfully initialised into the $\omega_c$ manifold the probability of shelving into the $\ket{\pm1/2}$ level is small as it necessitates two consecutive decays on the $\omega_b$ and $\omega_a$ transitions, both of which are considerably slower than our experiment/polarisation repetition rate (20~ms).
\end{enumerate}

We tried to improve the polarisation fidelity by incorporating direct driving on the $\omega_a$ transition using the method in Extended~Data~Fig.~7 during the polarisation protocol, thus leading to fast population exchange between $\ket{\pm1/2}$ and $\ket{\pm3/2}$. However, there was no improvement to the contrast of the resulting spin exchange oscillations thereby indicating that shelving into $\ket{\pm1/2}$ is not a limiting factor in our experiments.

\subsection{Analysis of Spin Exchange Dynamics}
In this section we present an analysis of the spin exchange dynamics on the $\omega_c$ register transition. The spin-exchange measurements in Fig.~2c are measured at a fixed ZenPol period of $2\tau=5.048~\mu$s leading to resonant interactions with the 991 kHz $\omega_c$ transition. However, analogous to the Rabi oscillations in a two-level system, the oscillation frequency and contrast of these spin transfer oscillations also depend on the detuning of the ZenPol sequence relative to the $^{51}$V transition. Specifically, we expect the following relations:
\begin{align}
J_\text{ex}(\delta)=\sqrt{J_\text{ex}(0)^2+\delta^2}\\
C(\delta)=\frac{J_\text{ex}(0)^2}{J_\text{ex}(0)^2+\delta^2}
\end{align}
Here $J_\text{ex}$ and $C$ are the spin-exchange frequency and oscillation contrast, respectively, and $\delta$ is the detuning of the ZenPol sequence resonance relative to a target nuclear spin transition. We polarise the register into $\ket{0_v}$ and measure the frequency detuning dependence of the spin-exchange oscillations in Extended~Data~Fig.~6c. These results agree well with the corresponding simulations shown in Extended~Data~Fig.~6b.

We also demonstrate control of the spin exchange frequency by varying the RF magnetic field amplitude ($B^\text{RF}$). Extended Data Figure~6d shows the spin-exchange dynamics for four different values of $B^\text{RF}$ = 0.8 G, 1.2 G, 1.6 G and 2.0 G. The inset in Fig.~2c plots extracted spin exchange frequencies $J_\text{ex}$ for a range of different $B^\text{RF}$ demonstrating linear dependence as expected and leading to accurate control of the engineered interaction strength (see main text for details).

\subsection{Single Excitation in $\omega_c$ Manifold}\label{singlespin}
The ability to shelve populations in different quadrupole levels enables the operation of the $^{51}$V register with an alternative set of many-body states: $\ket{0'_v}$ and $\ket{1'_v}$. For this experiment we polarise the $^{51}$V spins down the energy ladder on the $\omega_b$ and $\omega_c$ transitions leading to polarisation primarily into the $\ket{\pm3/2}$ level, with a small residual population in $\ket{\pm1/2}$. For the purpose of this analysis we will assume perfect polarisation into $\ket{\pm3/2}$, however we note that $\omega_a$ transition polarisation would be required for this.

We prepare the register $\ket{1'_v}$ state by injecting a single spin excitation on the $\omega_b$ transition (i.e. from $\ket{\pm3/2}\rightarrow\ket{\uparrow} = \ket{\pm5/2}$), this is achieved using the corresponding ZenPol resonance at $\omega_b$, $k=3$:
\begin{equation}
    \ket{1'_v}=\frac{1}{2}\left(\Ket{\uparrow,\frac{3}{2},\frac{3}{2},\frac{3}{2}}+\Ket{\frac{3}{2},\uparrow,\frac{3}{2},\frac{3}{2}}+\Ket{\frac{3}{2},\frac{3}{2},\uparrow,\frac{3}{2}}+\Ket{\frac{3}{2},\frac{3}{2},\frac{3}{2},\uparrow}\right)
\end{equation}
Here we omit the $\pm$ sign in the state label for simplicity. Subsequently, we prepare the $^{171}$Yb in $\ket{0_g}$ and induce a spin exchange oscillation between  $\ket{\uparrow}$ and $\ket{\downarrow} = \ket{\pm7/2}$ via a ZenPol sequence resonant with the $\omega_c$ transition. The resulting time evolution is given by

\begin{equation}
    \ket{\psi(t)} = \ket{0_g}\ket{1'_v}\cos\left(\frac{J'_\text{ex} t}{2}\right)-i\ket{1_g}\ket{0'_v} \sin\left(\frac{J'_\text{ex} t}{2}\right)
\end{equation}
where 
\begin{equation}
    \ket{0'_v}=\frac{1}{2}\left(\Ket{\downarrow,\frac{3}{2},\frac{3}{2},\frac{3}{2}}+\Ket{\frac{3}{2},\downarrow,\frac{3}{2},\frac{3}{2}}+\Ket{\frac{3}{2},\frac{3}{2},\downarrow,\frac{3}{2}}+\Ket{\frac{3}{2},\frac{3}{2},\frac{3}{2},\downarrow}\right)
\end{equation}
and $J'_\text{ex}=2b_{(5,\omega_c)}B^\text{RF}$. Notice that the spin-exchange oscillation rate, $J'_\text{ex}$, no longer has a $\sqrt{N}$ rate enhancement, this is because every ket in the $\ket{1'_v}$ and $\ket{0'_v}$ states contains only a single spin in the $\omega_c$-transition manifold. Using this manifold for information storage would have several benefits. For instance, direct microwave driving of the register $\omega_c$ transition would lead to Rabi oscillation between $\ket{0'_v}$ and $\ket{1'_v}$ and could therefore be used to realise local gates in this basis. Additionally, a second spin excitation is not allowed in this scheme, therefore the ZenPol sequence reproduces a complete two-qubit swap gate regardless of the $^{171}$Yb state. For these reasons, we believe that there may be some advantages to working with the $\{\ket{0'_v}$, $\ket{1'_v}\}$ manifold if the state initialisation fidelity into $\ket{\pm3/2}$ can be improved via direct $\omega_a$ transition polarisation. We leave this for future work.

\subsection{$T_2^*$ Coherence Discussion}
\label{coherence}
Here we provide detailed discussions regarding the \textsuperscript{51}V register coherence decay processes described in the main text. There are two magnetic interactions which limit the $T_2^*$ dephasing timescale: (1) the direct nuclear Zeeman interaction of each register spin with the Overhauser field (equation~\eqref{eq:Zeeman}) and (2) a contribution from the $^{171}$Yb Knight field \cite{Urbaszek2013}. In the latter case, the bath-induced $^{171}$Yb dipole moment generates a randomly fluctuating magnetic field at each $^{51}$V ion, the Knight field, which is described by
\begin{equation}
    \hat{H}_\text{nz,eff}=\mp g_{vz} \mu_N B^\text{OH}_{z}A_z\Hat{I}_z \label{eq:Knight}
\end{equation}
with
\begin{equation*}
    A_z=\frac{\mu_0\gamma_z^2(1-3n^2)}{8\pi r^3\omega_{01}}.
\end{equation*}
Here, the $-$ and $+$ cases in equation (\ref{eq:Knight}) correspond to $^{171}$Yb in $\ket{1_g}$ and $\ket{0_g}$, respectively. The constants are defined in Supplementary Information Section~\ref{hamderiv}. We note that $A_z$ corresponds to an effective local field amplification factor with value $A_z \approx 3.1$ for the register spins. We define the $^{171}$Yb Knight field to be $\mp A_z B^\text{OH}_{z}$. 

By applying periodic $\pi$ pulses to the $^{171}$Yb, we flip its state between $\ket{0_g}$ and $\ket{1_g}$, thereby switching the sign of the Knight field. This leads to the cancellation of $^{51}$V phase accumulation between successive free evolution periods, resulting in a longer coherence time. We numerically simulate the register coherence times using the method outlined in Supplementary Information Section~\ref{simulations}. When limited by the $^{171}$Yb Knight field, simulation yields a Gaussian decay with a $1/e$ coherence time of $33~\mu$s (equivalent to experimental results in Fig.~3a). We also predict an upper bound for the coherence time when decoupled from the $^{171}$Yb Knight field by turning off Hamiltonian terms associated with equation~\eqref{eq:Knight}, yielding an extended Gaussian decay of $417~\mu$s (equivalent to experimental results in Fig.~3b). These simulated values are consistent with the corresponding experimental results ($58\pm4~\mu$s and $225\pm9~\mu$s respectively) to within a factor of two. We note that this could indicate an error in our estimation of $g_{vz}$ by up to $25\%$, potentially caused by a small discrepancy in the position of the two \textsuperscript{51}V bath spins closest to \textsuperscript{171}Yb. Further analysis of these parameters is left for future work.

\subsection{$T_1$ Lifetime Discussion}
\label{T1decay}
We measure the population decay of both the $\ket{0_v}$ and $\ket{W_v}$ states (timescales $T_1^{(0)}$ and $T_1^{(W)}$ respectively) by preparing the $^{51}$V register in the appropriate state and waiting for a variable time, $t$, before swapping to the $^{171}$Yb for readout. 

The $\ket{0_v}$ state exhibits slow exponential decay with $1/e$ time constant $T_1^{(0)}=0.54\pm0.08$ s (Extended~Data~Fig.~8b). There are two contributions which could be limiting this decay:
\begin{enumerate}
\item
Resonant population exchange between the register spins and unpolarised frozen-core ‘dark spins’. For instance, the two nearest $^{51}$V ions (ions 1 and 2 in the table in Supplementary Information Section~\ref{localenv}) may interact resonantly with the neighbouring register spins. However, we cannot detect or polarise these dark spins since they only interact with the $^{171}$Yb via Ising-like $\hat{S}_z\hat{I}_z$ terms.
\item
Off-resonant population exchange between the register and detuned unpolarised bath spins.
\end{enumerate}

As for the $\ket{W_v}$ state, it exhibits a Gaussian decay with a much faster $1/e$ time constant of  $T_1^{(W)}=39.5\pm1.3~\mu$s (Extended~Data~Fig.~8a). This can be explained by considering the effect of {\it dephasing} on the register spins. Specifically, the $\ket{W_v}$ state which our $^{171}$Yb qubit interacts with is given as
\begin{equation*}
    \ket{W_v}=\frac{1}{2}\left(\Ket{\uparrow\downarrow\downarrow\downarrow}+\Ket{\downarrow\uparrow\downarrow\downarrow}+\Ket{\downarrow\downarrow\uparrow\downarrow}+\Ket{\downarrow\downarrow\downarrow\uparrow}\right).
\end{equation*}
Crucially, there are three additional orthogonal states required to span the $^{51}$V register single excitation subspace:
\begin{align*}
    \ket{\alpha_v}&=\frac{1}{2}\left(\Ket{\uparrow\downarrow\downarrow\downarrow}+\Ket{\downarrow\uparrow\downarrow\downarrow}-\Ket{\downarrow\downarrow\uparrow\downarrow}-\Ket{\downarrow\downarrow\downarrow\uparrow}\right)\\
    \ket{\beta_v}&=\frac{1}{2}\left(\Ket{\uparrow\downarrow\downarrow\downarrow}-\Ket{\downarrow\uparrow\downarrow\downarrow}+\Ket{\downarrow\downarrow\uparrow\downarrow}-\Ket{\downarrow\downarrow\downarrow\uparrow}\right)\\
    \ket{\gamma_v}&=\frac{1}{2}\left(\Ket{\uparrow\downarrow\downarrow\downarrow}-\Ket{\downarrow\uparrow\downarrow\downarrow}-\Ket{\downarrow\downarrow\uparrow\downarrow}+\Ket{\downarrow\downarrow\downarrow\uparrow}\right)
\end{align*}
We assume uncorrelated noise at each of the four $^{51}$V spins and apply a pure-dephasing master equation model. In the single excitation subspace, this becomes:
\begin{align}
    \Dot{\rho}=2\Gamma &\left[\mathcal{D}\left(\Ket{\uparrow\downarrow\downarrow\downarrow}\Bra{\uparrow\downarrow\downarrow\downarrow}\right)+\mathcal{D}\left(\Ket{\downarrow\uparrow\downarrow\downarrow}\Bra{\downarrow\uparrow\downarrow\downarrow}\right)\right.\\
&\left.+\mathcal{D}\left(\Ket{\downarrow\downarrow\uparrow\downarrow}\Bra{\downarrow\downarrow\uparrow\downarrow}\right)+\mathcal{D}\left(\Ket{\downarrow\downarrow\downarrow\uparrow}\Bra{\downarrow\downarrow\downarrow\uparrow}\right)\right] \rho
\end{align}
where the dephasing channel (Lindbladian) is given by 
\begin{equation}
    \mathcal{D}\left(\Hat{a}\right)\rho=\Hat{a}\rho\hat{a}^\dag-\frac{1}{2}\{\Hat{a}^\dag\Hat{a},\rho\}
\end{equation}
and $\Gamma$ is the dephasing rate on the $\omega_c$ transition of a single $^{51}$V spin. We solve this equation for different initial states $\rho(0)$. When $\rho(0) = \ket{0_v}\bra{0_v}$, dephasing does not contribute to $T_1^{(0)}$, i.e. $\rho(t)=\rho(0)$. However, when $\rho(0) = \ket{W_v}\bra{W_v}$ the state evolves according to
\begin{equation}
    \rho(t)=\ket{W_v}\bra{W_v}e^{-2\Gamma t}+\frac{1}{4}\left(1-e^{-2\Gamma t}\right)\mathbb{1}^{(\text{SEM})}
\end{equation}
where $\mathbb{1}^{(\text{SEM})}$ is the single excitation manifold identity operator:
\begin{equation*}
\mathbb{1}^{(\text{SEM})}=\ket{W_v}\bra{W_v}+\ket{\alpha_v}\bra{\alpha_v}+\ket{\beta_v}\bra{\beta_v}+\ket{\gamma_v}\bra{\gamma_v}
\end{equation*}
i.e. dephasing leads to decay of $\ket{W_v}$ into $\mathbb{1}^{(\text{SEM})}$ at rate $2\Gamma$. For completeness we also consider the decay of the off-diagonal coherence term $\rho_{01} = \bra{0_v}\rho \ket{W_v}$ and find that
\begin{equation}
    \rho_{01}(t)=\rho_{01}(0)e^{-\Gamma t}.
\end{equation}
Essentially, the pure dephasing model predicts $T_2^*=2T_1^{(W)}$ for our system.

We verify that dephasing is the main source of $\ket{W_v}$ population decay by demonstrating lifetime extension using the same motional narrowing approach employed to improve the coherence time (Supplementary Information Section~\ref{coherence}). Specifically, during the wait time, we apply a series of $\pi$ pulses to the $^{171}$Yb separated by $6~\mu$s leading to an extended lifetime of $T_1^{(W)}=127\pm8~\mu$s (Extended~Data~Fig.~8a). We note that both the bare and motionally-narrowed $T_1^{(W)}$ and $T_2^*$ times are close to the $T_2^*=2T_1^{(W)}$ limit identified above. We further extend the $T_1^{(W)}$ lifetime to $640\pm20~\mu$s using two $^{51}$V $\pi$ pulses applied during the wait time, thereby achieving dynamical decoupling from the nuclear Overhauser field (equivalent to the results in Fig.~3c).

Finally we note that if $T_1^{(W)}$ is limited by the \textsuperscript{171}Yb Knight field as a common noise source, there may be some discrepancy in the predictions of this model due to a high degree of noise correlation between the four $^{51}$V register spins arising from lattice symmetry. However, when performing motional narrowing we decouple the $^{171}$Yb Knight field and are likely limited by the, considerably less correlated, local Overhauser field. Further exploration of these correlated/uncorrelated fields is left for future work.

\subsection{Parity Oscillations and Coherence}
Here we derive an expression for the \textsuperscript{171}Yb--\textsuperscript{51}V Bell-state coherence $\rho_{01} = \bra{1_g0_v}\rho\ket{0_gW_v}$ in terms of the parity oscillation contrast with a correction factor. In particular, when reading out this coherence, we apply a $\sqrt{\text{swap}}$ gate which maps $\ket{\Psi^+} = \frac{1}{\sqrt{2}}(\ket{1_g0_v} - i\ket{0_gW_v})$ to $\ket{0_gW_v}$ and $\ket{\Psi^-}= \frac{1}{\sqrt{2}}(\ket{1_g0_v} + i\ket{0_gW_v})$ to $\ket{1_g0_v}$. Note that reading out the $^{171}$Yb state is sufficient to distinguish the $\ket{\Psi^+}$ and $\ket{\Psi^-}$ states in this measurement. We can account for the readout fidelity of the $\ket{\Psi^\pm}$ states by using a $\sqrt{\mathcal{F}_\text{sw,1}}$ factor (Methods), i.e. if the state $\ket{\Psi^+}$ ($\ket{\Psi^-}$) is perfectly prepared, $^{171}$Yb will be measured in state $\ket{0_g}$ ($\ket{1_g}$) with probability $\frac{1}{2}(1+\sqrt{\mathcal{F}_\text{sw,1}})$. To span the $^{171}$Yb--$^{51}$V Hilbert space, we also need to consider the effect of the readout $\sqrt{\text{swap}}$ gate when the system is initialised into the other two states: $\ket{1_gW_v}$ or $\ket{0_g0_v}$. To this end, we assign imperfect readout probabilities of $q_{11}$ and $q_{00}$ for $\ket{1_gW_v}$ and $\ket{0_g0_v}$, respectively. Specifically, we can represent the dependence of the parity readout on the input state using the following matrix relation:
\begin{equation}
    \begin{pmatrix}p_\text{1,Yb}\\p_\text{0,Yb}\end{pmatrix}=
    \mathcal{M}_\text{swap} \mathcal{M}_\text{wait} \begin{pmatrix}p_{11}\\p_{\Psi^+}\\p_{\Psi^-}\\p_{00}\end{pmatrix}
\end{equation}
with 
\begin{align}
    \mathcal{M}_\text{swap} &= \begin{pmatrix}q_{11}&&\frac{1}{2}(1-\sqrt{\mathcal{F}_\text{sw,1}})&&\frac{1}{2}(1+\sqrt{\mathcal{F}_\text{sw,1}})&&1-q_{00}\\1-q_{11}&&\frac{1}{2}(1+\sqrt{\mathcal{F}_\text{sw,1}})&&\frac{1}{2}(1-\sqrt{\mathcal{F}_\text{sw,1}})&&q_{00}\end{pmatrix}, \nonumber \\
    \mathcal{M}_\text{wait} &= \begin{pmatrix}1 && 0 &&0 && 0 \\
                        0 && \cos^2(\omega_c t/2) && \sin^2(\omega_c t/2) && 0 \\
                        0 && \sin^2(\omega_c t/2) && \cos^2(\omega_c t/2) && 0 \\
                        0 && 0 && 0 && 1 \end{pmatrix}. \nonumber
\end{align}
Here $p_\text{1,Yb}$ and $p_\text{0,Yb}$ are the probabilities of measuring the $^{171}$Yb qubit in $\ket{1_g}$ and $\ket{0_g}$, respectively, and $p_{\Psi^\pm} = \bra{\Psi^\pm}\rho\ket{\Psi^\pm}$ are the probabilities of being in the $\ket{\Psi^\pm}$ Bell states. The contrast $C_\text{parity}$ of the parity oscillation between $\ket{\Psi^+}$ and $\ket{\Psi^-}$ is extracted by measuring the difference in the $^{171}$Yb $\ket{1_g}$ populations measured at $t=0$ and $t=\pi/\omega_c$, allowing us to estimate the Bell state coherence as $|\rho_{01}| = C_\text{parity}/2\sqrt{\mathcal{F}_\text{sw,1}}$. This implies that uncorrected and corrected Bell state coherence values differ by a factor of $\sqrt{\mathcal{F}_\text{sw,1}}=0.72$. Using the results presented in  Fig.~4b we obtain corrected and uncorrected estimates for $|\rho_{01}|$ of $0.352\pm0.004$ and $0.254\pm0.003$ respectively.

\subsection{Bell State Fidelity Error Analysis}
To extract the Bell state fidelity and uncertainty, we perform a maximum likelihood analysis of the population and parity oscillation measurements, adopting a similar approach as in \cite{Bernien2013}. The population measurement involves a series of $n$ experiments with outcomes distributed between the four population states: $\{n_{00},n_{01},n_{10},n_{11}\}$ where $n=n_{00}+n_{01}+n_{10}+n_{11}$. The likelihood function for the uncorrected populations, $\{p_{00}, p_{01}, p_{10}, p_{11}\}$ has multinomial form:
\begin{equation} \label{Lp}
    \mathcal{L}\left(\{p_{ij}\}|\{n_{ij}\}\right)=\frac{n!}{n_{00}!n_{01}!n_{10}!n_{11}!}p_{00}^{n_{00}}p_{01}^{n_{01}}p_{10}^{n_{10}}p_{11}^{n_{11}}
\end{equation}
where we have assumed a prior uniform over the physical values of $\{p_{ij}\}$, i.e. $0\leq p_{ij}\leq 1$ and $\sum p_{ij}=1$. The likelihood function for the corrected populations, $\{c_{00}, c_{01}, c_{10}, c_{11}\}$, is obtained by substituting equation (11) into equation~\eqref{Lp} and assuming a prior uniform over the physical values of $\{c_{ij}\}$, i.e. $0\leq c_{ij} \leq1$ and $\sum c_{ij}=1$. Corrected populations are obtained by maximising this likelihood function. The error for a specific population (say $c_{00}$) is obtained by marginalising $\mathcal{L}\left(\{c_{ij}\}|\{n_{ij}\}\right)$ over the other three ($c_{01}, c_{10},c_{11}$) and taking a $68\%$ symmetric confidence interval.

We extract a likelihood function for the coherence by considering the following model:
\begin{equation}
    y_i=0.5+\sqrt{\mathcal{F}_\text{sw,1}}\rho_{01}\cos(\omega_c t_i)+\epsilon_i
\end{equation}
where $\{t_i,y_i\}$ are the parity oscillation data at the $i$\textsuperscript{th} point, $\rho_{01}$ is the corrected coherence, $\mathcal{F}_\text{sw,1}$ is the parity oscillation correction factor associated with the swap gate infidelity, and $\epsilon_i$ is the experimental error assumed to be normally distributed with $\mu=0$ and unknown $\sigma$. The likelihood function is given by
\begin{equation}
    \mathcal{L}\left(\rho_{01},\sigma|\{t_i,y_i\}\right)=\prod_i\frac{1}{\sqrt{2\pi}\sigma}\exp\left[-\frac{\left(y_i-0.5-\sqrt{\mathcal{F}_\text{sw,1}}\rho_{01}\cos(\omega_c t_i)\right)^2}{2\sigma^2}\right].
\end{equation}
We obtain a likelihood for the corrected coherence, $\mathcal{L}\left(\rho_{01}|\{t_i,y_i\}\right)$ by marginalising over $\sigma$.

The likelihood function for the fidelity is obtained by taking a product of the likelihood function for the populations with the likelihood function for the coherence and evaluating a contour integral at constant $\mathcal{F}$, given by 
\begin{equation}
    \mathcal{L}\left(\mathcal{F}\right)=\int_\mathcal{F}\mathcal{L}\left(\{c_{ij}\}|\{n_{ij}\}\right)\mathcal{L}\left(\rho_{01}|\{t_i,y_i\}\right)d\rho_{01}\prod_{ij}dc_{ij}.
\end{equation}
The Bell state fidelity is extracted by maximising this likelihood and the error is evaluated as a symmetric $68\%$ confidence interval.

\section*{Acknowledgements}
This work was funded by the Institute of Quantum Information and Matter, an NSF Physics Frontiers Center (PHY-1733907) with support from the Moore Foundation, NSF 1820790, Office of Naval Research Award No. N00014-19-1-2182, Air Force Office of Scientific Research Grant No. FA9550-18-1-0374 and No. FA9550-21-1-0055, Northrop Grumman, General Atomics, and Weston Havens Foundation. The device nanofabrication was performed in the Kavli Nanoscience Institute at the California Institute of Technology.  J.R. acknowledges the support from the Natural Sciences and Engineering Research Council of Canada (NSERC) (PGSD3-502844-2017). A.R. acknowledges the support from the Eddleman Graduate Fellowship. J.C. acknowledges support from the IQIM postdoctoral fellowship. 
We thank J. Kindem, J. G. Bartholomew, N. Yao, A. Sipahigil, M. Lei and T. Xie for useful discussion, and M. Shaw for help with superconducting photon detectors.

\end{document}